\documentclass[fleqn,usenatbib]{mnras}

\usepackage[T1]{fontenc}

\DeclareRobustCommand{\VAN}[3]{#2}
\let\VANthebibliography\thebibliography
\def\thebibliography{\DeclareRobustCommand{\VAN}[3]{##3}\VANthebibliography}

\usepackage{graphicx}	
\usepackage{amsmath}	
\usepackage{amssymb}	
\usepackage{subfig}
\usepackage{booktabs}
\usepackage{afterpage}
\usepackage{float}
\usepackage{tablefootnote}
\usepackage{threeparttable}

\usepackage{newtxtext,newtxmath}

\newcommand{\angstrom}{\mbox{\normalfont\AA}}
\newcommand{\sgr}{$s_{gr}$}
\newcommand{\sbv}{$s_{BV}$}

\newcommand{\nickel}{$^{56}$Ni}
\newcommand{\mni}{M_\mathrm{Ni56}}
\newcommand{\msol}{M_\odot}
\newcommand{\fg}{2003fg-like SNe}
\newcommand{\es}{2002es-like SNe}

\title[\fg\ $\gamma$-ray deposition]{High $\gamma$-ray escape time in 2003fg-like supernovae: A challenge to proposed models}

\author[Sharon \& Kushnir]{
	Amir Sharon$^{1}$\thanks{E-mail: amir.sharon@weizmann.ac.il},
	  Doron Kushnir$^{1}$,
        and Eden Schinasi-Lemberg$^{1,2}$
	\\
	$^{1}$Dept.of Particle Phys. \& Astrophys., Weizmann Institute of Science, Rehovot 76100, Israel\\
        $^{2}$Dept. of Physics, NRCN, Beer-Sheva 84190, Israel\\
}

\date{Accepted XXX. Received YYY; in original form ZZZ}

\pubyear{2021}

\begin{document}
\label{firstpage}
\pagerange{\pageref{firstpage}--\pageref{lastpage}}
\maketitle

\begin{abstract}
A rare subclass of Type Ia supernovae (SNe Ia), named after the prototype SN 2003fg, includes some of the brightest SNe Ia, often called "super Chandrasekhar-mass" SNe Ia. We calculate the $\gamma$-ray deposition histories and the \nickel\ mass synthesized in the explosion, $\mni$, for eight \fg. Our findings reveal that the $\gamma$-ray escape time, $t_0$, for these objects is $ t_0\approx45\text{--}60 \,$ days, significantly higher than that of normal SNe Ia. \fg\ are distinct from normal SNe Ia in the $ t_0 $--$ \mni $ plane, with a noticeable gap between the two populations. The observed position of \fg\ in this plane poses a significant challenge for theoretical explosion models. We demonstrate that the merger of two white dwarfs (WDs) and a single star exceeding the Chandrasekhar limit fail to reproduce the observed $ t_0 $--$ \mni $ distribution. However, preliminary calculations of head-on collisions of massive WDs show agreement with the observed $ t_0 $--$ \mni $ distribution.
\end{abstract}





\section{Introduction}
\label{sec:intro} 

Type Ia supernovae (SNe Ia) are believed to result from thermonuclear explosions of white dwarfs (WDs), but their progenitor systems and explosion mechanisms remain subjects of debate \citep[for a review, see, e.g.,][]{Maoz2014}. Despite exhibiting continuous and relatively uniform properties that make them useful as extragalactic standard candles, SNe Ia display a wide range of luminosities \citep{Phillips1993}. Their peak bolometric luminosities range is $\sim10^{42}\text{--}10^{43}\,\text{erg}\,\text{s}^{-1}$, and the synthesized \nickel\ masses, $\mni$, which power these events through its radioactive decay chain, vary between $ \sim0.1\text{--}1\msol $ \citep[see][for a recent compliation]{Sharon2020}. Moreover, several SNe Ia sub-types have been identified, each with distinct spectroscopic characteristics and varying luminosity levels. These include the faint 1991bg-like SNe, 2002cx-like SNe, and 2002es-like, the bright 1991T-like SNe, and others \citep[see][for a review]{Ashall2020}.

One of the rarest classes of SNe Ia, named after the prototype SN 2003fg \citep{Howell2006}, are among the brightest SNe Ia and are sometimes referred to as "super Chandrasekhar-mass" SNe Ia. The over-luminous light curves of the first few identified objects in this class suggested $\mni$ exceeding the Chandrasekhar mass limit, $M_\text{ch}\approx1.4\,M_{\odot}$. However, additional objects later classified into this group require less extreme \nickel\ masses. Apart from their high luminosities, SN 2003fg and similar events exhibit diverse properties not consistently shared across all SNe of this type. Nonetheless, several characteristics are common to all SN 2003fg-like SNe Ia: broad light curves, a slow expansion velocity gradient before maximum light, and a very strong $\lambda$6580 C II absorption feature that persists well past the $B$-band maximum \citep{Ashall2021}. The near-infrared (NIR) light curves of SN 2003fg-like SNe Ia show significant differences from typical SNe Ia. They are notably brighter and broader, and the $Y$ and $H$ bands lack a secondary maximum. This behavior is also observed in some optical bands, where the light curves in the $r$ and $i$ bands do not display a second maximum or a 'shoulder' as seen in other types of SNe Ia.


\begin{figure*}
	\includegraphics[width=\textwidth]{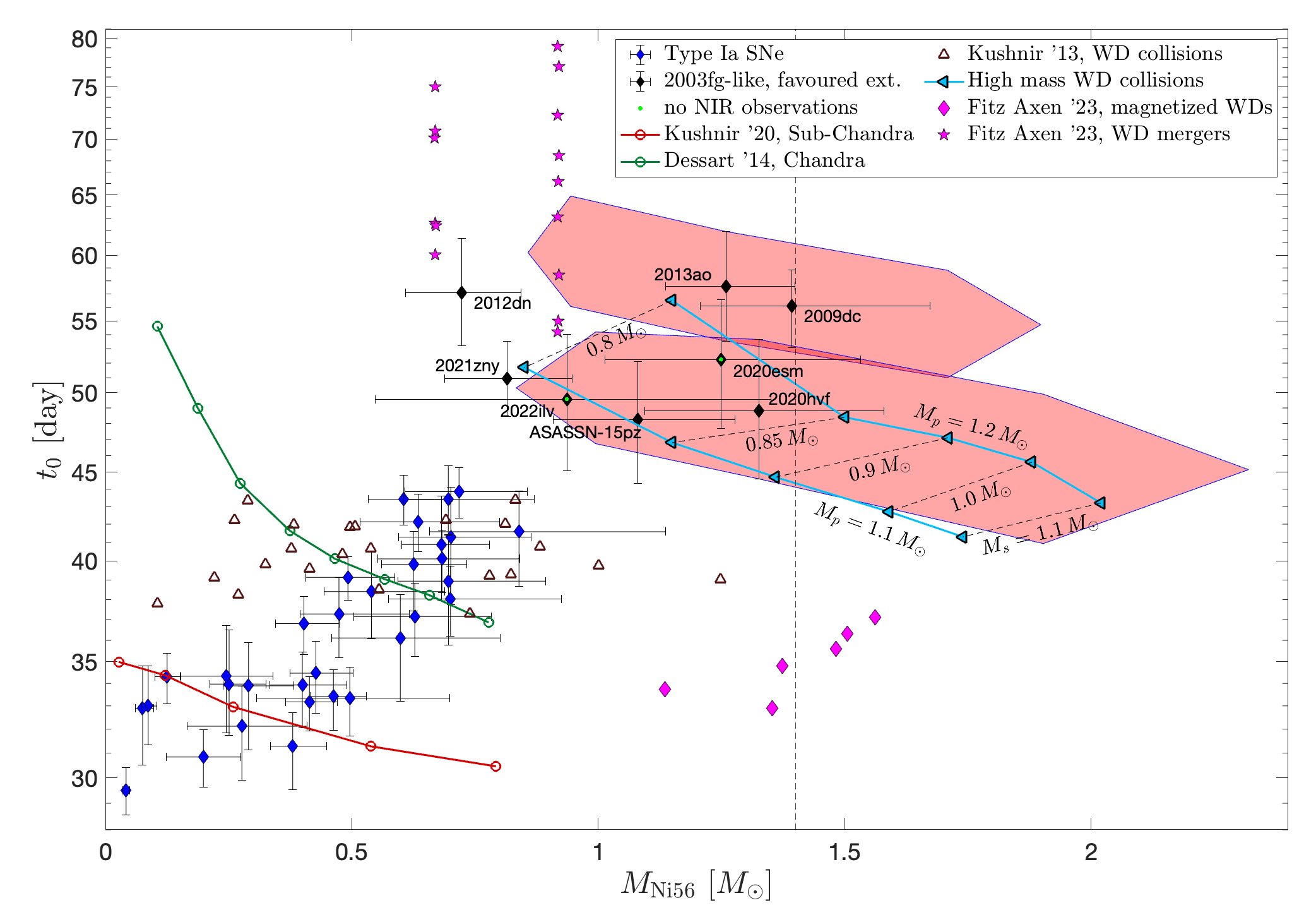}
	\caption{$\gamma$-ray escape time, $t_0$, as a function of the synthesized \nickel\ mass, $\mni$. Black diamonds represent the \fg\ sample with the preferred host-extinction value. Red-shaded regions demonstrate the ranges of $t_0$ and $\mni$ values for SN 2020hvf and SN 2013ao, resulting from uncertainties in the host extinction correction. Green dots represent objects without NIR observations. Blue symbols indicate normal SNe Ia from the sample of \protect\citet{Sharon2020}. \fg\ are distinct from normal SNe Ia in the $t_0$--$\mni$ plane, with an apparent gap between these two populations. Normal SNe Ia models, including sub-Chandra detonation \protect\citep[][red line]{Kushnir2020}, Chandrasekhar-mass explosions \protect\citep[][green line]{Dessart2014}, and WD head-on collisions \protect\citep[][upward-facing, brown triangles]{Kushnir2013}, generally exhibit $t_0$ values that are too low to match the observed \fg\ values. The explosion models of \protect\citet{FitzAxen2023}, representing two suggested progenitor channels for \fg -- the merger of two WDs (magenta stars) and a single star exceeding the Chandrasekhar limit (magenta diamonds) -- fail to reproduce the observed $t_0$--$\mni$ distribution. Head-on collisions of massive WDs (left-facing blue triangles), featuring an ONe WD primary with a mass of $M_P=1.1$ or $1.2,M_\odot$, and a CO secondary with masses ranging from $M_S=0.8\text{--}1.1\,M_\odot$ (configurations with the same primary (secondary) mass are connected by solid (dashed) lines, with the corresponding mass indicated next to each line), show agreement with the observed $t_0$--$\mni$ distribution for $M_S=0.8\text{--}0.85\,M_\odot$.}
	\label{fig:t0-ni}
\end{figure*}

The origin of 2003fg-like events is even more ambiguous than normal SNe Ia. Due to their higher luminosities and unique characteristics, it is unclear whether normal SNe Ia and \fg\ originate from the same progenitor systems or result from different explosion channels. Several models have been proposed for \fg, including interaction with a dense circumstellar medium \citep{Hachinger2012}, and the explosion of a WD surpassing the Chandrasekhar limit, made possible by the merger of two WDs, rapid rotation, or high magnetic fields \citep{Yoon2005,Das2013,FitzAxen2023}. Violent WD mergers are also considered viable candidates \citep{Hicken2007,Kwok2023}. An accurately measured SNe Ia luminosity function (LF), which describes the intrinsic luminosity distribution of these SNe, can help constrain the progenitor systems. For non-\fg\ Ia, the LF has been consistent with a single progenitor channel \citep{Sharon2021}, although previous works support multiple progenitor channels \citep{Pakmor2013,Ashall2016,Hakobyan2020}. However, a survey dedicated to the LF of \fg\ has not yet been conducted, and their rarity and unknown completeness factors make such an attempt challenging. Recently, \cite{Desai2023} calculated the rate of \fg\ to be $30^{+20}_{-20} \,\text{yr}^{-1} \,\text{Gpc}^{-3}\, h^3_{70}$, though their sample consists of only two objects\footnote{The normal SNe Ia rate is $(2\text{--}3)\times10^4\,\text{yr}^{-1} \,\text{Gpc}^{-3}\, h^3_{70}$ \citep{Sharon2021,Desai2023}}.



In this paper, we constrain the progenitor system of \fg\ by studying their $\gamma$-ray deposition histories. For a small enough $\gamma$-ray optical depth, each $\gamma$-ray photon has a small probability of interacting with matter, such that the deposition function is proportional to the column density, which scales as $t^{-2}$, where $t$ is the time since the explosion. The $ \gamma $-ray escape time, $ t_0 $, is defined by \citep{Jeffery1999}:
\begin{equation}\label{eq:dep_late}
f_\text{dep}(t) = \frac{t_0^2}{t^2},\;\;\;f_\text{dep}\ll 1,
\end{equation}
where $f_\text{dep}(t)$ is the $\gamma$-ray deposition function, indicating the fraction of generated $\gamma$-ray energy deposited in the ejecta. Along with $\mni$, $t_0$ can be accurately measured from bolometric light curves. The measurement requires an extensive coverage of the SNe, in both time and wavelength range, limiting the number of qualifying SNe. The method requires observations at sufficiently late times, when the ejecta becomes optically thin, and the luminosity equals the instantaneous deposition:
\begin{equation}\label{eq:L_eq_Q}
    L(t)=Q_\text{dep}(t)= Q_\gamma(t) f_\text{dep}(t)+Q_\text{pos}(t),
\end{equation}
where $Q_\text{dep}(t)$ is the energy deposited in the ejecta from radioactive decay, $Q_\gamma(t)$ ($Q_\text{pos}(t)$) is the radioactive energy generated from $\gamma$-ray photons (kinetic energy of positrons). Applying this method to SNe Ia, \cite{Sharon2020} found $\gamma$-ray escape times of $t_0\approx30\text{--}45\,\text{day}$. The obtained $ t_0 $--$ \mni $ distribution was inconsistent with all known SNe Ia explosion models \citep[][see also Figure~\ref{fig:t0-ni}]{Sharon2020b}, where a precise calculation of $t_0$ from an ejecta model can be performed easily, without requiring radiation transfer calculations \citep[][]{Wygoda2019}.  






Here, we employ the methods of \cite{Sharon2020} to measure $\mni$ and $t_0$ for \fg, aiming to constrain their progenitor system. Due to their rarity, our sample size is limited to eight SNe: four SNe from the CSP \fg\ sample of \citep[][SN 2009dc, SN 2012dn, SN 2013ao, and ASSASN-15pz]{Ashall2021} and four additional objects (SN2020esm \citep{Dimitriadis2022}, SN 2020hvf \citep{Jiang2021}, SN 2021zny \citep{Dimitriadis2023}, and SN 2022ilv \citep{Srivastav2023}). The main results are presented in Figure~\ref{fig:t0-ni}, where we compare our derived values of $ t_0 $ and $ \mni $ for \fg\ with those of SNe Ia. As can be seen in the figure, the range of $ t_0 $ values of \fg\ is $ t_0\approx45$--$60\,\text{day} $, higher than those of normal SNe Ia. Additionally, their $\mni$ values are generally higher as well. \fg\ are distinct from normal SNe Ia in the $ t_0 $--$ \mni $ plane, with an apparent gap between these two populations. However, the small sample size prevents us from conclusively determining whether they constitute two separate populations.

Figure~\ref{fig:t0-ni} also compares the observed sample with the one-dimensional explosion models of \cite{FitzAxen2023}, which represent two suggested progenitor channels for \fg: the non-violent merger of two WDs and a single star exceeding the Chandrasekhar limit. 
The figure shows that the single star explosions fail to reproduce the observed $ t_0 $--$ \mni $ distribution, and the non-violent mergers only partly fall within the observed range. We discuss the reasons for this discrepancy and outline the requirements that explosion models must meet to match the observed $ t_0 $--$ \mni $ distribution. One model that meets these requirements is head-on collisions of massive WDs, for which preliminary calculations (cyan lines) show agreement with the observed $ t_0 $--$ \mni $ distribution.

The paper is structured as follows: Section~\ref{sec:t0-ni} provides a brief overview of our methods and calculates the $ \gamma $-ray deposition histories of the SNe in our sample. In Section~\ref{sec:models}, we compare our findings with explosion models from the literature. We discuss the implications of our results and provide conclusions in Section~\ref{sec:conclusions}. In Appendix ~\ref{sec:colors}, we analyze the host extinction of the SNe by examining their colors at the peak brightness of each band and estimating a range of possible extinction values. Appendix~\ref{sec:2002es} investigates a sample of three \es\ and explores potential connections with \fg, following recent claims for such a connection based on early UV color evolution \citep{Hoogendam2023}.

\section{The $ \lowercase{t}_0 - M_{\text{N\lowercase{i}56}} $ relation of 2003\lowercase{fg}-like SN\lowercase{e}}	
\label{sec:t0-ni}

In this section, we construct the bolometric light curves of our \fg\ sample, using their published photometry and the methods outlined in \cite{Sharon2020}, and calculate their position in the $ t_0 $--$ \mni $ plane.

The process starts by constructing light curves in all available bands, employing interpolations and extrapolations to fill in any missing epochs. These light curves are converted to flux densities at their effective wavelengths to create a spectral energy distribution (SED) for each epoch. The SED is estimated using a blackbody fit for wavelengths longer than the longest effective wavelength band. These longer wavelengths contribute less than a few percent to the total flux when $JHK$ measurements are available, which applies to most of the SNe in our sample. Exceptions include SN 2020esm and SN 2022ilv, which lack redward observations beyond the $z$ band, rendering their values less precise than those of the rest of the sample. We repeat this process for the range of host extinction values discussed in Appendix~\ref{sec:colors}. The resulting bolometric light curves for the favored extinction values are shown in Appendix~\ref{app:deposition plots}, and are included in the supplementary material.

Following the construction of the bolometric light curves, we apply the methods from \cite{Wygoda2019} and \cite{Sharon2020} to calculate $ t_0 $ and $\mni$ for our sample. This calculation relies on the Katz integral \citep{Katz2013}, given by:
\begin{equation}\label{eq:integral}
\begin{split}
&QT = LT-ET,\\
&QT \equiv \int_0^t Q_\text{dep}(t')t'dt',\:\:\:&LT \equiv \int_0^t L(t')t'dt',
\end{split}
\end{equation}
where $Q_\text{dep}(t)$ is defined in Equation~\eqref{eq:L_eq_Q}, $L(t)$ is the bolometric luminosity, and $ET$ represents the integrated time-weighted luminosity that would be emitted if no $^{56}$Ni was produced. As in the case of the SNe Ia sample in \cite{Sharon2020}, we assume $ ET=0 $, indicating that our model does not account for other energy sources besides \nickel\ decay. To describe the deposition fraction over time, we use the following interpolating function, which connects the expected behavior at early and late times:
\begin{equation}\label{eq:deposition}
f_\text{dep}(t)=\frac{1}{\left(1+\left(t/t_0\right)^n\right)^{\frac{2}{n}}},
\end{equation}
where $n$ is a parameter that controls the smoothness of the interpolation and is determined during the fitting process. Combining Equation~\eqref{eq:integral} (assuming $ET=0$) with Equation~\eqref{eq:L_eq_Q}, yields the following relation:
\begin{equation}\label{eq:integral_ratio}
\frac{L(t)}{LT(t)}=\frac{Q_\text{dep}(t)}{QT(t)}.
\end{equation}
The advantage of using the Katz integral is that the left-hand side of this equation is a distance-independent observational quantity. Furthermore, the right-hand side, which represents the theoretical calculations, is independent of $\mni$, thus removing the degeneracy between $t_0$ and $\mni$ in the fit \citep[see][for further details]{Wygoda2019}.

We fit Equation \eqref{eq:integral_ratio} by minimising the expression
\begin{equation}\label{eq:likelihood}
\frac{N_\text{bins}}{N_\text{obs}}\sum_{t_i\in t_{L=Q}} \left[\left(\frac{L(t_i)}{LT(t_i)}-\frac{Q_{\text{dep}}(t_i)}{QT(t_i)}\right)\frac{LT(t_i)}{L_{\text{err}}(t_i)}\right]^2,
\end{equation}
where $L_\text{err} $ is the luminosity error, $N_\text{obs}$ is the number of observations, and $N_\text{bins}$ is the number of independent time bins. The number of bins is defined as the number of times that $Q_\text{dep}$ changes by $10\%$ over the time range of each SN. The ratio $N_\text{bins}/N_\text{obs}$ affects only the uncertainty of the parameters and not the best-fit values. The time range $t_{L=Q}=[t_\text{min},t_\text{max}]$ includes the periods when the assumption  $L=Q_\text{dep}$ is valid. The upper limit, $t_\text{max}$, is set by the latest epoch where the observations follow the deposition model. The lower limit, $t_\text{min}$, is the earliest epoch at which the observations within $t_{L=Q}$ uniformly scatter around the best fit. As mentioned above, Equation~\eqref{eq:likelihood} is independent of $\mni$ and the distance, so the fit is performed over $t_0$ and $n$ alone. $\mni$ is subsequently determined by comparing the luminosity in the fitted range to the deposited radioactive energy. 

The uncertainty of the parameters is estimated using a Markov Chain Monte Carlo (MCMC) algorithm, implemented with the MCMCSTAT Matlab package\footnote{https://mjlaine.github.io/mcmcstat/}. The likelihood function is given by Equation~\eqref{eq:likelihood}, and the priors are uniformly distributed over reasonable domains.

The inferred values of $t_0$ and $\mni$ are presented in Table~\ref{tab:results} and illustrated in Figure~\ref{fig:t0-ni}. The best-fit deposition models for each object are depicted as solid black lines in Figure~\ref{fig:2009dc}, which should align with observations within $t_{L=Q}$. The radioactive energy generation rates (corresponding to $f_\text{dep}=1$) are shown as dashed black lines in the left panels of the figure. The obtained $ t_0 $ and $ \mni $ values for the preferred host-extinction values are indicated as black diamonds in Figure~\ref{fig:t0-ni}. The ranges of $ t_0 $ and $ \mni $ values, resulting from uncertainties in the host-extinction correction, are demonstrated for SN 2020hvf and SN 2013ao as red-shaded regions. Additionally, normal SNe Ia (blue symbols) from the sample of \citet{Sharon2020} are included\footnote{The sample has been supplemented with several new objects. Furthermore, several SNe have been excluded from the sample upon reevaluation, as their light curves did not have complete coverage of the peak, necessary to compute the Katz integral accurately.}. 

Figure~\ref{fig:t0-ni} reveals that \fg\ differ from normal SNe Ia regarding their $\mni$ and $ t_0 $ values. The \nickel\ masses for \fg\ are generally much higher, reaching up to $ 1.4\,M_\odot $ for the preferred extinction values. There is some overlap in the \nickel\ masses due to the low value of SN 2012dn, which might be attributed to an unusually early formation of dust in the ejecta \citep{Taubenberger2019}. The escape times of all \fg\ are significantly longer than those of SNe Ia, with values of $ \approx45$--$60 \,\text{day}$. In the $ t_0 $--$ \mni $ plane, \fg\ are distinct from normal SNe Ia, with a noticeable gap between these two populations. However, the small sample size limits our ability to conclude whether they form two separate populations definitively.

\begin{table*}
	\vspace{-0.25 cm}
	\begin{threeparttable}
		\caption{Parameters of the \fg\ sample and the results of the \nickel\ deposition model. The derived parameters of the deposition model are the median values of the posterior distribution, together with the $68\%$ confidence levels.}
		\renewcommand{\arraystretch}{1.3}
		\begin{tabular}{lccccccc}\midrule
			Name        &  $ \mu $\tnote{a}  & $ E(B-V)_\text{MW} $\tnote{b}  &$ E(B-V)_\text{host,fav} $\tnote{c} &  \sbv\tnote{d} & $M_{\text{Ni}56}\:[M_\odot]$ & $t_0$ [day] & Source\tnote{e}   \\[-0.15cm]
                2009dc     &  34.85$\,\pm\,$0.08  &  0.07 &   0.0 &  1.30& $ 1.39^{+0.28}_{-0.19}$ & $ 56^{+ 3}_{- 3}$ & \cite{Ashall2021}\\ 
 2012dn     &  33.15$\,\pm\,$0.15  &  0.06 &   0.1 &  1.01& $ 0.72^{+0.12}_{-0.11}$ & $ 57^{+ 4}_{- 4}$ & \cite{Ashall2021}\\ 
 2013ao     &  36.39$\,\pm\,$0.04  &  0.03 &   0.1 &  1.05& $ 1.26^{+0.14}_{-0.12}$ & $ 58^{+ 4}_{- 4}$ & \cite{Ashall2021}\\ 
 ASASSN-15pz&  33.85$\,\pm\,$0.12  &  0.02 &   0.0 &  1.39& $ 1.08^{+0.20}_{-0.17}$ & $ 48^{+ 4}_{- 4}$ & \cite{Ashall2021}\\ 
 2020esm    &  35.98$\,\pm\,$0.15  &  0.02 &   0.0 &  1.16& $ 1.25^{+0.28}_{-0.24}$ & $ 52^{+ 4}_{- 5}$ & \cite{Dimitriadis2022}\\ 
 2020hvf    &  32.45$\,\pm\,$0.15  &  0.04 &   0.1 &  1.14& $ 1.33^{+0.25}_{-0.23}$ & $ 49^{+ 5}_{- 4}$ & \cite{Jiang2021}\\ 
 2021zny    &  35.14$\,\pm\,$0.15  &  0.04 &   0.0 &  1.25& $ 0.82^{+0.13}_{-0.13}$ & $ 51^{+ 3}_{- 2}$ & \cite{Dimitriadis2023}\\ 
 2022ilv    &  35.28$\,\pm\,$0.44  &  0.11 &   0.0 &  1.02& $ 0.94^{+0.41}_{-0.39}$ & $ 50^{+ 4}_{- 4}$ & \cite{Srivastav2023}\\ 
 \midrule
 \end{tabular}
		\begin{tablenotes}
			\item [a] Distance modulus
			\item [b] Galactic extinction towards the SN
			\item [c] The favored value for the host extinction. 
			\item [d] Color stretch parameter for the $B$ and $V$ bands.
			\item [e] Source for the photometry and distance.
		\end{tablenotes}
		\label{tab:results}
	\end{threeparttable}
\end{table*}

\section{Comparison to models}
\label{sec:models}

This section compares the observed $ t_0 $--$ \mni $ distribution of \fg\ to theoretical models. Due to the limited number of simulations specifically addressing the explosion of \fg, we also consider models for normal SNe Ia. The $\mni$ values of the models are taken from the original publications, and we calculate the $t_0$ values using $\gamma$-ray MC simulations following the method outlined in \citep{Sharon2020} \footnote{We are grateful to the authors of the original publications for providing the explosion ejecta profiles.}. The results are presented in Figure~\ref{fig:t0-ni}. Below, we provide a description of each model.

The normal SNe Ia models include sub-Chandra detonation \citep[][red line]{Kushnir2020}, Chandrasekhar-mass explosions \citep[][green line]{Dessart2014}, and WD head-on collisions \citep[][upward-facing, brown triangles]{Kushnir2013}. While not aiming to explain \fg, the Chandra and sub-Chandra models have ejecta profiles with $\mni$ that overlap with the observed low-end \nickel\ masses of \fg. However, these models have $t_0$ ${\approx}30\text{--}37\,\text{day}$, significantly lower than the any of the observations. The collision models also overlap with the low-end \nickel\ masses and have slightly higher $t_0$ values, reaching up to ${\approx}44\,\text{day}$, but these are still insufficient to match the observed values. 

We next examine the \fg\ models of \citet{FitzAxen2023}, which represent two suggested progenitor channels for \fg: a merger of two WDs with a total mass exceeding $M_\text{ch}$, where the secondary WD disrupts and accretes onto the primary WD before explosion, and an explosion of a WD with strong magnetic fields that support mass above the Chandrasekhar limit. The merger models consist of a primary WD with masses $M_P=1.09$ or $1.18\,M_\odot$, and a disrupted secondary leading to total system masses of $1.8$, $2$ or $2.2\,M_\odot$. The initial profiles of these merger models are one-dimensional (1D), where the secondary was isotropically distributed around the primary. The magnetized WD models were constructed with varying central densities and magnetic field strengths, resulting in WD masses of ${\approx}1.4\text{--}1.75\,M_\odot$. Both sets of models were ignited at the center to initiate the explosion, and nucleosynthesis and hydrodynamics calculations were performed until the ejecta reached homologous expansion.

In the merger models, only the primary WD successfully ignites, while the secondary remains either unburned or only partially burned into intermediate-mass elements. These model yield $\mni{\approx}0.67\,M_\odot$ ($0.92\,M_\odot$) for $M_P=1.09\,M_\odot$ ($1.18\,M_\odot$). Applying our $\gamma$-ray MC simulations, the merger models have $t_0\approx54\text{--}79\,\text{day}$. Since only the primary WD successfully ignites, the explosion of this scenario has about the same $\mni$ and kinetic energy as that of a single, $M_P$ mass WD explosion, but with roughly twice the ejecta mass. The obtained $t_0$ values can be understood by noting that it scales with the square root of the column density (Equation~\eqref{eq:dep_late}):
\begin{equation}
    \label{eq:t0_scale}
    t_0\propto\sqrt{\int \rho dv} \propto \sqrt{\frac{M_\text{ej}}{v^2}}\propto\frac{M_\text{ej}}{\sqrt{E_\text{kin}}},
\end{equation}
where $M_\text{ej}$, $v$, and $E_\text{kin}$ are the ejecta mass, characteristic velocity, and kinetic energy, respectively. Therefore, the escape times of these models are expected to be about twice that of a single WD explosion with the same primary mass. Indeed, sub-Chandra detonations of WDs with $M\approx1.1\,M_\odot$ have $t_0\approx30\,\text{day}$ (see Figure~\ref{fig:t0-ni}). While some WD merger profiles have $t_0$ values compatible with \fg , their generally low $\mni$ is only compatible with the lowest $\mni$ members of \fg. Increasing the primary mass would raise their $\mni$ yield, possibly allowing better agreement with the observed $\mni$ distribution (see Section~\ref{sec:conclusions}). Additionally, these models are 1D, despite the asymmetrical nature of a merger effect, and this approximation might significantly alter the results. Recent polarimetry measurements suggest highly aspherical explosion in these events \citep{Nagao2024}.


The results of the magnetized models differ significantly from the merger models, with lower $\gamma$-ray escape times of $t_0\approx33\text{--}37 \,\text{day}$ and higher \nickel\ masses, $\mni\approx1.13\text{--}1.56\,M_\odot$. Although the \nickel\ yields can match the observed range, the escape times of these models deviate significantly from the observations. These results are also consistent with their high ejecta velocities, exceeding the observed Si II velocities of \fg\ \citep{FitzAxen2023}. Given the significant discrepancy in escape times, it is unlikely that these models are the progenitors of \fg. In Section~\ref{sec:conclusions}, we propose potential modifications to these models that could increase $t_0$ while preserving the $\mni$ values. 

We supplement these models with head-on collisions of high-mass WDs, shown as left-facing blue triangles, which will be detailed in a future paper. These models feature an ONe WD primary with a mass of $M_P=1.1$ or $1.2\,M_\odot$, and a CO secondary with masses ranging from $M_S=0.8\text{--}1.1\,M_\odot$. Configurations with the same primary (secondary) mass are connected by solid (dashed) lines, with the corresponding mass indicated next to each line. The $\mni$ synthesized in these models matches the large amounts of $\mni$ observed in \fg. Additionally, the $t_0$ values fall within the observed range for a subset of the configurations. This subset is characterized by large mass differences, with the secondary masses at the lower end of the mass range, $M_S=0.8\text{--}0.85\,M_\odot$ (calculations with lower secondary masses are currently unavailable, and will be published in future work). The longer escape times of this subset occur because the detonation wave initiated by the collision fails to ignite the entire ONe primary star, leaving behind a bound remnant of ${\approx}0.1\text{--}0.4\,M_\odot$. The incomplete burning and the reduced nuclear energy due to the initial ONe composition result in significantly lower kinetic energy of the ejecta and increased $\gamma$-ray escape times. When the mass difference is smaller, or the primary is a CO WD, incomplete burning does not occur. 




\section{Discussion}
\label{sec:conclusions}	

In this study, we derived the observed $ t_0 $--$ \mni $ distribution for \fg. Unlike the typical photometric properties that are usually examined, this quantity relates to the late-time light curve evolution and requires extended observation periods ( $ \gtrsim 80 $ days from the explosion) to be accurately determined. The advantage of using this relation is its ease of computation from SN ejecta profiles, allowing for reliable and robust constraints on different SNe explosion models. Our findings, shown in Figure~\ref{fig:t0-ni}, indicate that the $t_0$ range for \fg\ is $ t_0\approx45$--$60\,\text{day} $, which is higher than that of normal SNe Ia. Additionally, their $\mni$ values are generally higher. \fg\ are distinct from normal SNe Ia in the $ t_0 $--$ \mni $ plane, with a noticeable gap between these two populations. However, the small sample size prevents us from conclusively determining whether they constitute two separate populations.

The observed position of \fg\ in the $ t_0 $--$ \mni $ plane presents a significant challenge for theoretical explosion models, as most models we have considered fail to replicate both the high $t_0$ values and the large amounts of $\mni$ observed in these events (see Figure~\ref{fig:t0-ni}). The difficulty arises because producing a large amount of $\mni$ requires the progenitor system to have a large mass, approximately $M_\text{ch}$, at a high density at the moment of ignition. This condition conflicts with the merger models of \citet{FitzAxen2023} and likely rules out all non-violent merger models with primary mass not approaching the Chandrasekhar limit as viable progenitors for \fg. 

Focusing on progenitor systems with sufficient high-density mass at ignition, the main observational challenge is the high $t_0$ values. The magnetized WD models of \citet{FitzAxen2023} show that $t_0$ values from WDs surpassing the Chandrasekhar limit are generally too short. This issue might be resolved by considering an initial composition with heavier elements, which would reduce the available nuclear energy and, thus, the kinetic energy of the explosion. Another potential solution is accumulating low-density mass on top of the WD, increasing the total ejecta mass while maintaining the same kinetic energy (equivalent to increasing the primary mass of the non-violent merger models toward the Chandrasekhar limit). Both options warrant further study.

Another type of progenitor system with sufficient high-density mass at ignition involves two undisrupted (massive enough) WDs. If both WDs are entirely burned in the explosion, the resulting $t_0$ will generally be too low. This is because the $t_0$ for the two WDs should be similar to that of each WD separately, which is akin to the value for normal SNe Ia. Consequently, collisions and violent mergers of CO WDs, where both stars are entirely burned, cannot explain \fg . This argument suggests that for a system with two undisrupted (massive enough) WDs, at least one must be only partially burned, or another mechanism must be in place to reduce the kinetic energy of the ejecta. Interestingly, observations support a large fraction of unburned material in such scenarios \cite[see, e.g.,][]{Ashall2021,Siebert2023}.

An example of such a model involves collisions between a $1.1\text{--}1.2\,M_\odot$ ONe WD and a $0.8\text{--}0.85$ CO WD. In these configurations, the massive and highly dense ONe WD does not undergo complete burning because the temperature near the point of contact does not rise sufficiently to initiate nuclear burning. As a result, up to ${\approx}0.4\,M_\odot$ of unburned material remains. The incomplete burning and reduced nuclear energy due to the initial ONe composition led to the significantly lower kinetic energy of the ejecta and increased $\gamma$-ray escape times. Discovery of the unburned remnant would strongly support this model and could be achieved using similar methods as those applied to the Type Iax SN 2008ha \citep{Foley2014}. Violent WD mergers might also achieve incomplete burning of the primary or secondary WD, but we currently lack models to test this hypothesis.

\fg\ remain one of the most enigmatic extragalactic transients. Unveiling the nature of these objects is particularly challenging due to their puzzling and diverse properties and their extreme rarity. We have demonstrated that some models of \fg\ can be ruled out, but our sample remains limited. Discovering additional objects could help reveal the full range of $t_0$ and $\mni$ for these SNe and place further constraints on their origin.

\section*{Acknowledgements}
We thank Luc Dessart and Margot Fitz-Axen for sharing their ejecta profiles. We thank Boaz Katz for useful discussions. DK is supported by a research grant from The Abramson Family Center for Young Scientists, and by the Minerva Stiftung.

\section*{Data Availability}
The data underlying this article are available in the article and in its online supplementary material.

\bibliographystyle{mnras}
\bibliography{bibliography} 

\appendix	

\section{Intrinsic colors and extinction}
\label{sec:colors}
To derive the intrinsic bolometric light curves of the SNe, accounting for line-of-sight extinction from the host galaxy is essential. However, accurately determining this quantity poses challenges. Normal SNe Ia display a 'blue edge' in their color curves, and their host extinction can be estimated relatively accurately by matching the observed colors to the intrinsic, extinction-free colors of SNe Ia  \citep[see, e.g.,][]{Tripp1998,Phillips2012}. The intrinsic colors have been previously characterized using the color stretch parameter \sbv\ \citep[][]{Burns2014,Burns2018}, defined as the time difference between the $B$-band peak time and the $(B-V)$ color peak time:
\begin{equation}
    \label{eq:sbv}
    s_{BV}=\frac{t_{\text{max},(B-V)}-t_{\text{max},B}}{30\,\text{day}}.
\end{equation}
Moreover, \sbv\ has been found to correlate strongly with the SNe Ia luminosity \citep{Burns2018,Sharon2021}. Estimating the extinction of \fg\ using this approach presents several challenges. Firstly, the stretch parameter alone cannot reliably distinguish \fg\ from normal SNe Ia. Although \sbv\ values for \fg\ tend to be higher than those for normal SNe Ia, some values overlap with the higher end of the normal SNe Ia distribution (e.g., $ s_{BV}\approx1.06 $ for SN 2013ao, a typical value for normal SNe Ia). Secondly, the light curves of \fg\ are considerably less uniform, raising doubts about the existence of uniform intrinsic colors among them. Determining such uniformities would be challenging due to the limited number of observed objects.

Hence, rather than precisely determining the host extinction for these objects, our approach is to establish the possible range of host extinction values, assuming they are related to SNe Ia. Following a methodology akin to \cite{Burns2018}, we analyze the peak magnitude difference between two selected bands as a function of the stretch parameter \sbv. Peak magnitudes and \sbv\ for the SNe are derived from observed photometry using Gaussian process interpolations. For SNe lacking sufficient $B$ and $V$ band data, adequate measurements in the $g$ and $r$ bands enable computation of the \sgr\ parameter, analogous to \sbv\, but using Sloan filters. Subsequently, \sbv\ is calculated from \sgr\ using the relation established in \cite{Ashall2020}.

\begin{figure*}
	\includegraphics[width=\textwidth]{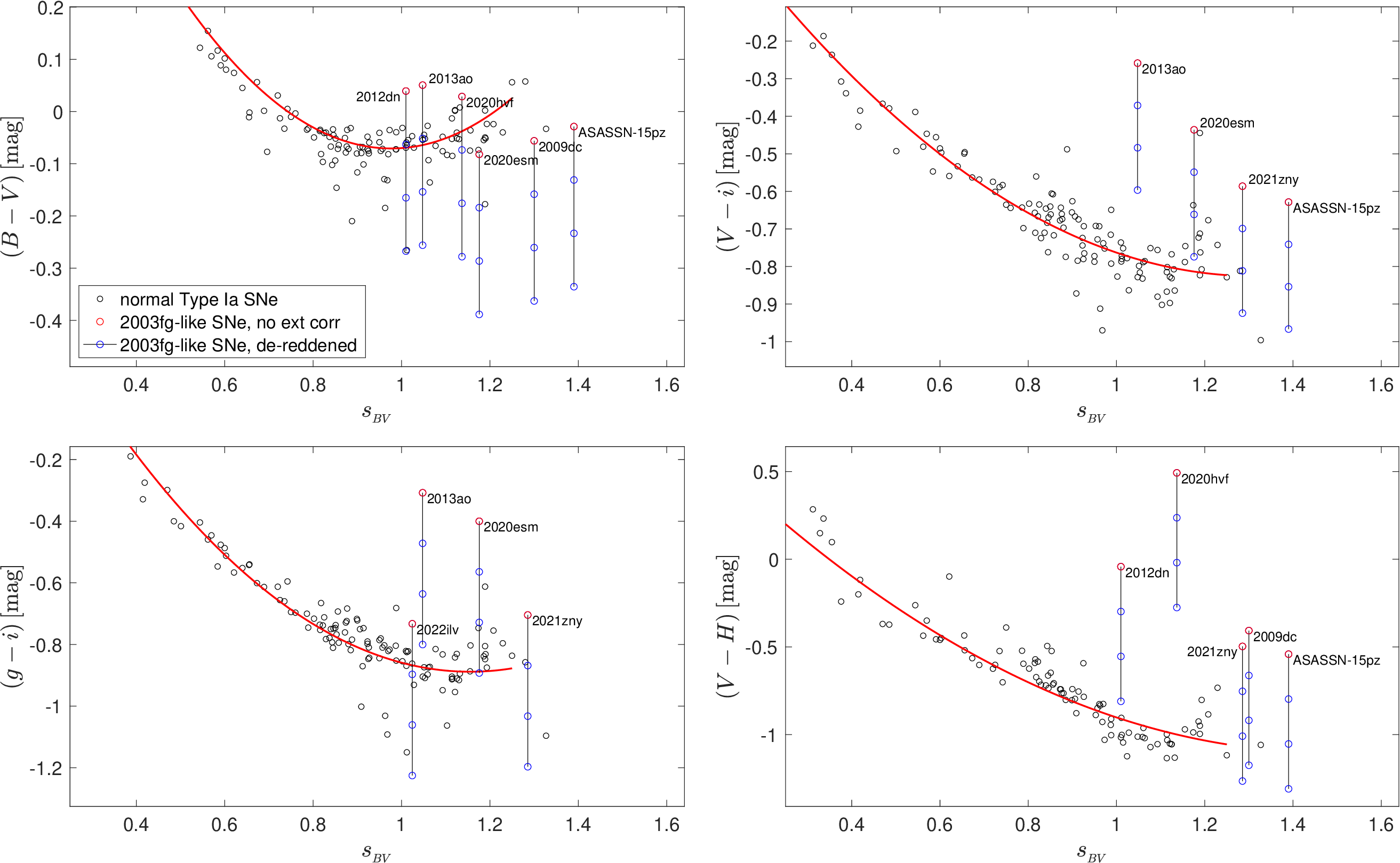}
	\caption{Peak magnitude difference of several filters as a function of $ s_{BV} $. Normal SNe Ia from the CSP sample of \protect\cite{Burns2018}, corrected for both host and galactic extinction, are represented by black symbols. The red line shows the 'blue edge' from that work, indicating intrinsic SNe Ia colors. \fg\ are depicted as red and blue symbols, where red symbols indicate values corrected for galactic extinction only, while blue symbols represent additional host corrections with $E(B-V)=0.1$, $0.2$, and $0.3\,\text{mag}$.
}
	\label{fig:colors}
\end{figure*}

Figure~\ref{fig:colors} shows the peak magnitudes difference of SNe Ia and \fg\ for several pairs of bands as a function of \sbv. Black symbols indicate normal SNe Ia, corrected for host and galactic extinction. The \sbv, peak magnitudes, and host extinction values are taken from \cite{Burns2018}. The red symbols mark \fg\ with only galactic extinction correction. The solid red lines mark the blue edge of \cite{Burns2018}, calculated using the relation between \sbv\ and the peak magnitudes given in that work. The peak colors of the \fg\ seem to form two groups - one with high \sbv\ values and relatively similar colors containing SN 2009dc, ASASSN-15pz and SN2021zny, and one with typical or moderately high SNe Ia \sbv\ values containing SN 2012dn, SN 2013ao, SN 2020esm, SN 2020hvf and SN2022ilv. The peak colors of the latter group are diverse and, except for SN 2022ilv, are redder than the other \fg.
As can be seen in the figure, the $(B-V)$ color of the \fg\ sample is near the extinction-corrected values of SNe Ia, while colors involving filters with longer wavelengths have large differences between the two SNe types. This strengthens the claim that extinction alone cannot account for all color differences between SNe Ia and \fg. In contrast to the rest of the sample, the $(B-V)$ colors of SN 2012dn, SN 2013ao, and SN 2020hvf lie somewhat above the SNe Ia blue edge. We, therefore, test the effects of extinction (for our entire \fg\ sample) by de-reddening the colors according to $E(B-V)$ values of up to $0.3\,\text{mag}$, assuming $R_V=3.1$. The modified colors are shown as blue circles, extending blue-wards from the unmodified values, corresponding to $0.1\,\text{mag}$ increments in the $E(B-V)$ de-reddening. While only ${\approx}0.1\,\text{mag}$ of $E(B-V)$ is required to match the $B-V$ color of 2012dn, SN 2013ao, and SN 2020hvf to SNe Ia, matching other colors requires significantly larger amounts of extinction, which would cause the $(B-V)$ colors to be much bluer than any SNe Ia. 



We, therefore, assume that the \fg\ cannot significantly exceed the SNe Ia $(B-V)$ blue edge but instead are allowed to be redder, resulting in low values of host extinction for the SNe in our sample. This is supported by the typical environments in which these SNe are found, consisting mostly of low-luminosity dwarf galaxies \citep{Ashall2021}. For SNe with other type of hosts, the explosion was far away from the host's center (ASSASN-15pz, SN 2020esm, SN 2020hvf, and SN 2021zny), or no host galaxy was found (SN 2022ilv). Other works have found low or zero values of host extinction as well \citep{Chen2019,Ashall2021,Jiang2021,Dimitriadis2022,Dimitriadis2023}. 

Therefore, we choose a 'favored' extinction value for SN 2012dn, SN 2013ao, and SN 2020hvf as the value that matches the SNe $ B-V $ color to the blue edge, found to be $ 0.1\,\text{mag} $ for all three SNe. For the remaining SNe, we assume zero extinction as the favored value, except for SN 2009dc, for which we adopt $ E(B-V)=0.1\,\text{mag} $, as estimated by \cite{Chen2019}. The stretch and extinction values for the SNe are listed in Table~\ref{tab:results}. We assume an uncertainty of $0.1\,\text{mag}$ for the host extinction. 

We perform an additional analysis to further constrain the host extinction and verify our results. We compute the bolometric light curves of our sample for $E(B-V)$ values between 0 and $0.5\,\text{mag}$ and apply our fitting procedure from Section~\ref{sec:t0-ni} for each light curve. We then assess the quality of the fit by examining the normalized deviation of the observations from the best-fit model using the equation
\begin{equation}
    \label{eq:fit_validity}
    \frac{L/LT}{Q_\text{dep}/QT}-1,
\end{equation}
easily derived from Equation~\eqref{eq:integral_ratio}. If the underlying model properly describes the observations, this quantity should be centered around zero for all epochs within $t_{L=Q}$. We find that, for most SNe, this is not the case for light curves corrected with high extinction values, where significant trends in the normalized deviations as a function of time are observed.

\begin{figure}
    \centering
    \includegraphics[width=\columnwidth]{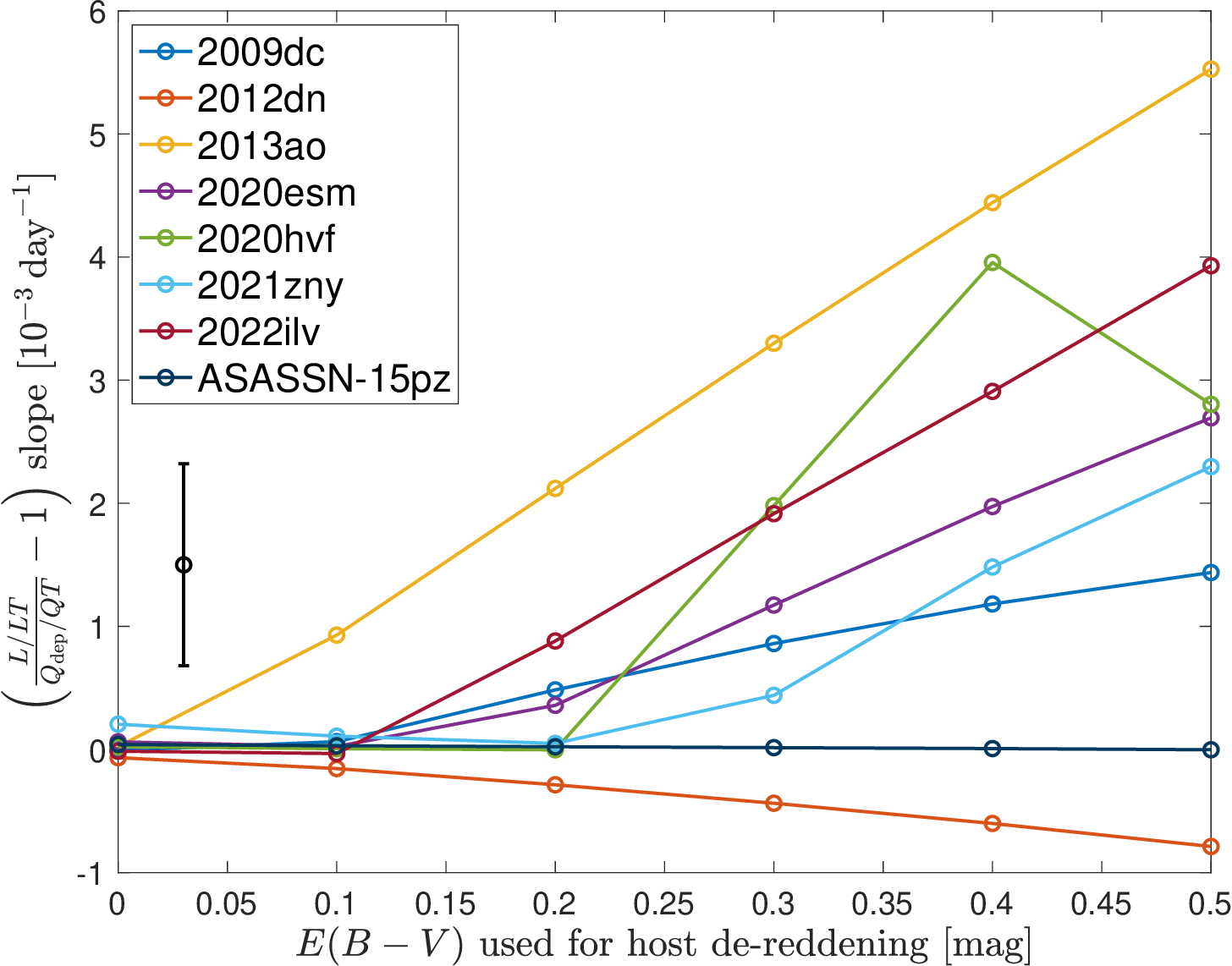}
    \caption{Slope of the normalized deviation (Equation~\eqref{eq:fit_validity}) as a function of the host $E(B-V)$, used to correct the bolometric light curve. The mean error of the slope is indicated in the middle left of the figure. For the majority of the sample, the slope increases for $E(B-V)>0.1\,\text{mag}$, indicating that the model poorly describes the observations at higher extinction values.}
    \label{fig:slope_fits}
\end{figure}

To quantify this trend, we calculate the slope of the normalized deviations over time. The results, shown in Figure \ref{fig:slope_fits}, plot the slope of the normalized deviation against the $E(B-V)$ value used to correct for host extinction. For the majority of the sample, the slope increases for $E(B-V)>0.1\,\text{mag}$, indicating the model poorly describes the observations at higher extinction values. Except for ASASSN-15pz, the absolute value of the slope at $E(B-V) = 0.5\,\text{mag}$ is $\gtrsim0.001\,\text{day}^{-1}$. Given that the time range $t_{L=Q}$ of the SNe in our sample is typically ${\approx}50$ days, this implies a systematic shift in the normalized deviations of at least ${\approx}5$ percent. Therefore, the SNe in our sample are less likely to experience high levels of host extinction.

\section{Relation to 2002\lowercase{es}-like SNe}
\label{sec:2002es}

2002es-like SNe are a rare, subluminous class of SNe Ia, with spectra similar to 1991bg-like SNe but with significantly wider light curves compared to other faint SNe Ia \citep{Taubenberger2017}. Despite substantial luminosity differences, they share several characteristics with \fg, such as the absence of a secondary maximum in their $I/i$-band light curves and the presence of carbon absorption features around peak light \citep{Cao2015,Li2023}. A recent study by \cite{Hoogendam2023} examined and compared the early UV light curves of both these types of SNe. They found that both classes can be distinguished from other subtypes of SNe Ia by their early UV color evolution and have members with identifiable bumps in their rising light curves, a feature not observed in other subtypes of SNe Ia. Here, we explore whether the $t_0$--$\mni$ distribution of these SNe suggests a similar connection between the two subtypes.

Due to the rarity and faint luminosities of \es, there is only one object, SN 2016ije, with late-time measurements and NIR observations. We collected other \es\ with sufficient late-time measurements required for our analysis and attempted to correct for their missing NIR flux. The SNe and their photometry, distance and extinction sources are: SN 2002es \citep{Ganeshalingam2012}, SN 2006bt \citep{Foley2010}, iPTF14atg \citep{Cao2015}, SN 2019yvq \citep{Miller2020}, and SN 2022vqz \citep{Xi2023}.

\begin{figure}
    \centering
    \includegraphics[width=\columnwidth]{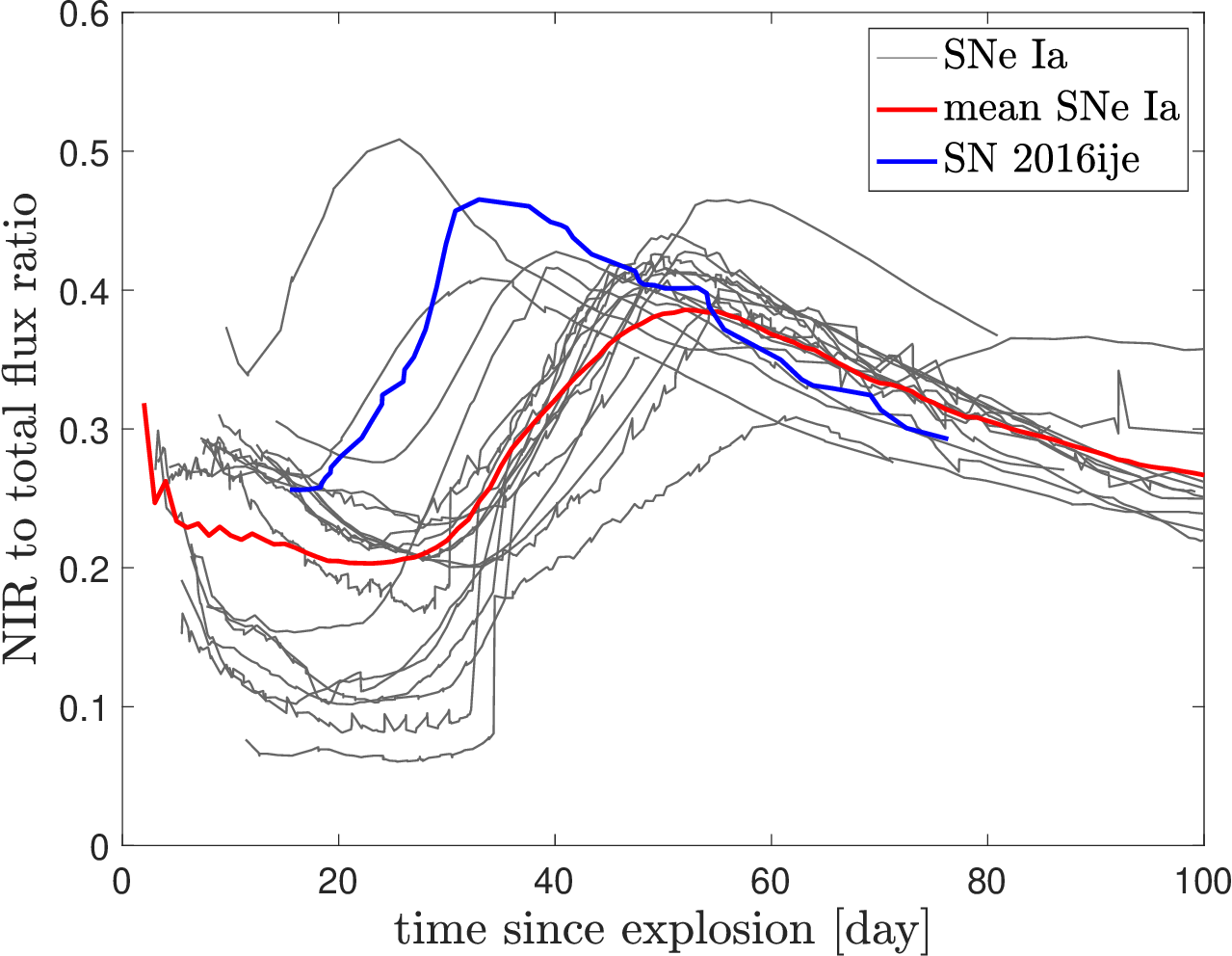}
    \caption{Evolution of the ratio of NIR flux ($\lambda>8000\,\angstrom$) to the total flux as a function of time since explosion. Gray lines represent the SNe Ia sample, the red line denotes the sample's mean value, and the blue line illustrates SN 2016ije. SN 2016ije displays elevated values and faster evolution of the NIR flux fraction compared to typical SNe Ia, resembling characteristics of low-luminosity SNe Ia.}
    \label{fig:IRfrac}
\end{figure}

We construct the bolometric light curves of these SNe following the methods described in Section~\ref{sec:t0-ni}. However, for all SNe except SN 2016ije, the SED is integrated up to the $I/i$-band effective wavelength, and the flux at longer wavelengths is estimated using a correction function. Given the limited number of \es\ with NIR observations, it is unclear whether the NIR flux fraction is uniform across different SNe or varies significantly. Therefore, we apply two different corrections: the first uses the mean NIR flux fraction of the SNe Ia in our sample, and the second uses the NIR flux fraction of SN 2016ije. These corrections are illustrated in Figure~\ref{fig:IRfrac}. The gray lines and red line represent the evolution and mean value of the NIR flux fraction of the SNe Ia sample, respectively. The blue line shows the NIR flux fraction of SN 2016ije, which exhibits higher values and faster evolution than most SNe Ia, similar to low-luminosity SNe Ia.

As outlined in Section~\ref{sec:t0-ni}, we computed the $t_0$ and $\mni$ for the \es\ of our sample. Among those lacking NIR observations, only iPTF14atg and SN 2019yvq exhibited a satisfactory fit to our model, while SN 2002es, SN 2006bt, and SN 2022vqz showed subpar fits. In particular, SN 2002es displayed an inadequate fit to the \nickel\ decay model due to a rapid decline of its light curve from ${\approx}30$ days after peak. This behavior was also noted by \cite{Ganeshalingam2012}, who proposed several hypotheses to explain the fast decline, including dust formation, an IR catastrophe, or alternative powering mechanisms during peak brightness. However, \cite{Ganeshalingam2012} discussed the challenges associated with each explanation, highlighting that the fading of SN 2002es is currently poorly understood. Our model provides a better fit for the light curves of SN 2006bt and SN 2022vqz, but not sufficiently well. It is unclear whether their inadequate fit is due to a similar cause as observed in SN 2002es or is influenced by the quality of their observational data.

The NIR corrections for SN 2019yvq and iPTF14atg significantly affect the results, with differences of up to ${\approx}30$ percent in both $t_0$ and $\mni$ between the two correction methods. For both SNe, employing the correction derived from SN 2016ije results in a poorer fit to our model. Thus, we favor the results obtained using the mean correction of SNe Ia. The bolometric light curves and best-fit deposition models for each object are given in Table~\ref{tab:results_02es} and shown in Figure~\ref{fig:fits_2002es}.

\begin{figure*}
    \centering
    \includegraphics[width=\textwidth]{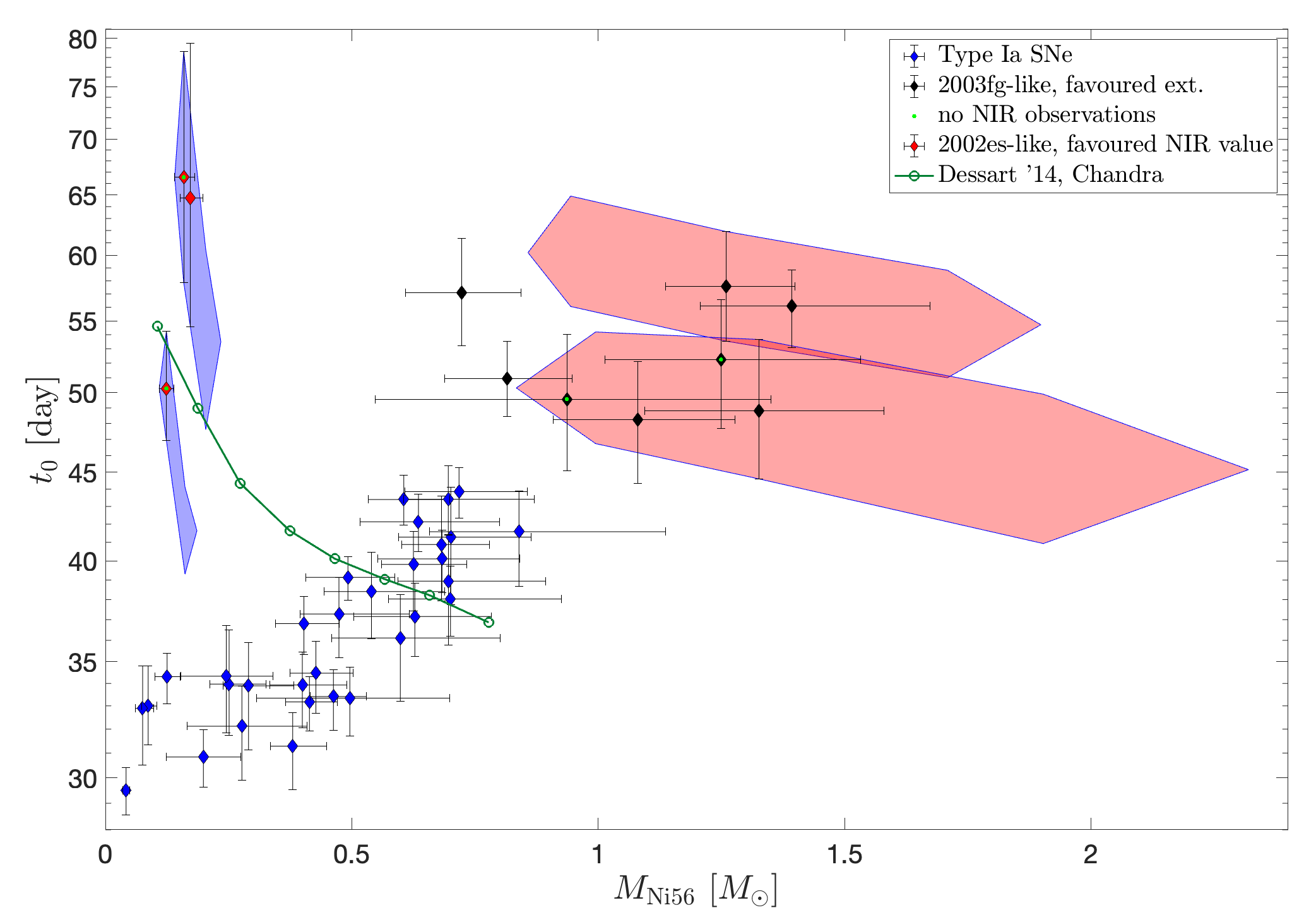}
    \caption{Same as Figure~\ref{fig:t0-ni}, supplemented with observations of three \es\ (iPTF14atg, SN 2016ije and SN 2019yvq). Models are not shown except for the Chandrasekhar-mass explosion models of \protect\cite{Dessart2014}. The red symbols represent SN 2016ije and the favored values of SN 2019yvq and iPTF14atg. The blue-shaded regions around the favored values of SN 2019yvq and iPTF14atg (additionally marked with a green dot) indicate results due to variations in the NIR correction function. All \es\ are distinctly separated from the rest of the SNe Ia. Similar to \fg, the \es\ in our sample are characterized by high $t_0$ values. However, a significant gap exists between the observed $\mni$ of the two types.}
    \label{fig:t0Ni02es}    
\end{figure*}

The $t_0$--$\mni$ distribution of the sample of \es\ is shown in Figure~\ref{fig:t0Ni02es}, along with the normal SNe Ia and \fg\ samples. SN 2016ije and the favored values of SN 2019yvq and iPTF14atg are displayed as red symbols. The blue-shaded region denotes the range of results due to variations in the NIR correction applied to SN 2019yvq and iPTF14atg (additionally marked with a green dot). The sample of \es\ have $\gamma$-ray escape times of $t_0\approx40\text{--}60\,$d, and $\mni\approx0.1\text{--}0.2\,M_\odot$. These results suffer from large uncertainties, particularly due to the absence of NIR measurements for most of our sample. Nonetheless, they clearly separated from the distribution of normal SNe Ia, showing considerably higher $t_0$ values compared to normal SNe Ia with similar $\mni$ values across the entire range of NIR corrections. This behavior is consistent with their positions in the $B$-band peak magnitude and decline rate distribution as reported by \cite{Taubenberger2017}, indicating broader light curves compared to SNe Ia with similar peak magnitudes. However, the $B$-band decline rates are somewhat lower than the average for normal SNe Ia, while their $\gamma$-ray escape times are notably longer.

We supplement Figure \ref{fig:t0Ni02es} with the Chandrasekhar-mass explosion models of \citet[][green line]{Dessart2014}. As seen in the figure, the low-luminosity configurations of this model are in rough agreement with the observed \es, a result of the large ejecta mass of these models.


\begin{table*}
    \centering
    \caption{Parameters of the \es\ sample and the results of the \nickel\ deposition model. The derived parameters of the deposition model are the median values of the posterior distribution, together with the $68\%$ confidence levels.}
    \begin{threeparttable}
    \renewcommand{\arraystretch}{1.3}
    \begin{tabular}{llcccccc}
			Name   & NIR treatment \tnote{a}   &  $ \mu $\tnote{b}  & $ E(B-V)_\text{MW} $\tnote{c}  &$ E(B-V)_\text{host} $\tnote{d} & $M_{\text{Ni}56}$ & $t_0$  & Source\tnote{e}  \\[-0.15cm]
			& & & & &$ (M_\odot) $ &(d)\\\midrule 
 iPTF14atg   & SNe Ia mean&  34.86$\,\pm\,$0.07  &  0.01 &  0.00& $   0.16^{+  0.02}_{-  0.02}$ & $   67^{+  12}_{-   9}$ &  \cite{Cao2015} \\ 
 iPTF14atg   & SN 2016ije &  34.86$\,\pm\,$0.07  &  0.01 &  0.00& $   0.20^{+  0.03}_{-  0.03}$ & $   54^{+   7}_{-   6}$ &  \\ 
 2016ije     & $JHK$ obs &  35.68$\,\pm\,$0.05  &  0.05 &  0.00& $   0.17^{+  0.03}_{-  0.02}$ & $   65^{+  15}_{-  10}$ &  \cite{Miller2020}\\ 
 2019yvq     & SNe Ia mean&  33.14$\,\pm\,$0.11  &  0.02 &  0.03& $   0.12^{+  0.02}_{-  0.01}$ & $   50^{+   4}_{-   3}$ & \cite{Xi2023} \\ 
 2019yvq     & SN 2016ije &  33.14$\,\pm\,$0.11  &  0.02 &  0.03& $   0.16^{+  0.02}_{-  0.02}$ & $   42^{+   2}_{-   2}$ &  \\ 
    \end{tabular}
    \begin{tablenotes}
			\item [a] Method for calculation of the NIR flux. Options are: (1) Object has $JHK$ measurements. (2) Use of mean SNe Ia NIR flux fraction. (3) Use of SN 2016ije NIR flux fraction.
			\item [b] Distance modulus.
			\item [c] Galactic extinction towards the SN.
			\item [d] Host extinction.
			\item [e] Source for the photometry and distance.
		\end{tablenotes}
	\end{threeparttable}
 \label{tab:results_02es}
\end{table*}

The connection of \es\ to \fg\ also seems to hold, to some degree, in our analysis. Similar to \fg, the \es\ in our sample are characterized by high $t_0$ values. However, a significant gap exists between the observed $\mni$ of the two types. The \nickel\ masses falling within this gap correspond with typical SNe Ia. This gap is currently devoid of any objects, as no SNe Ia with a typical $\mni$ was observed with such high $t_0$ values.

\section{Bolometric light curves and the results of the fitting method for the entire sample}
\label{app:deposition plots}

We present the best-fit bolometric light curves and the fitting results for our \fg\ sample (Figure~\ref{fig:2009dc}, using the favored extinction values) and \es\ sample (Figure~\ref{fig:fits_2002es}, using the favored NIR values). The bolometric light curves are also included in the online supplementary material.

\begin{figure*}
	\includegraphics[width=0.9\textwidth]{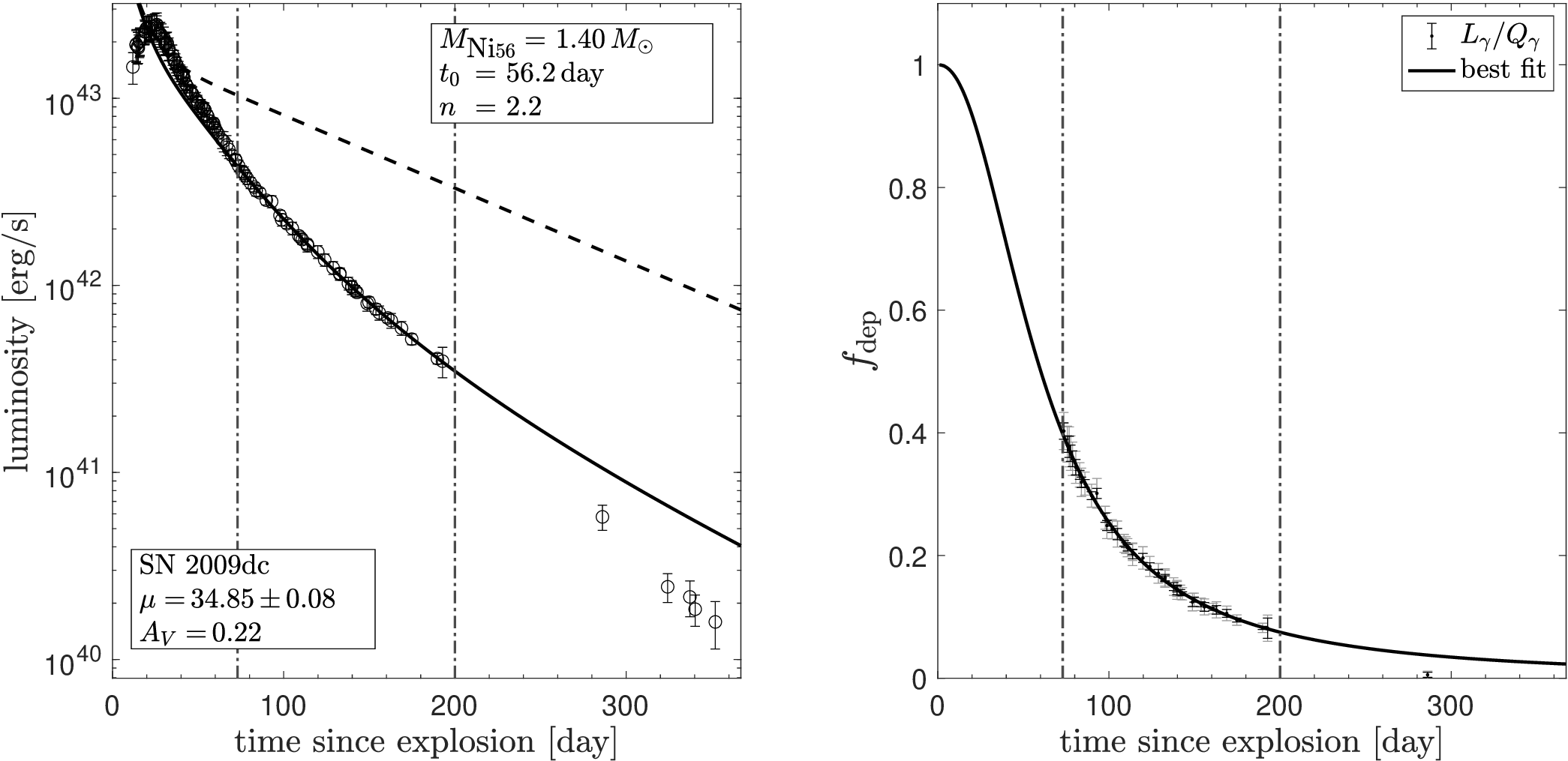}
	
	\vspace{1 cm}
	\includegraphics[width=0.9\textwidth]{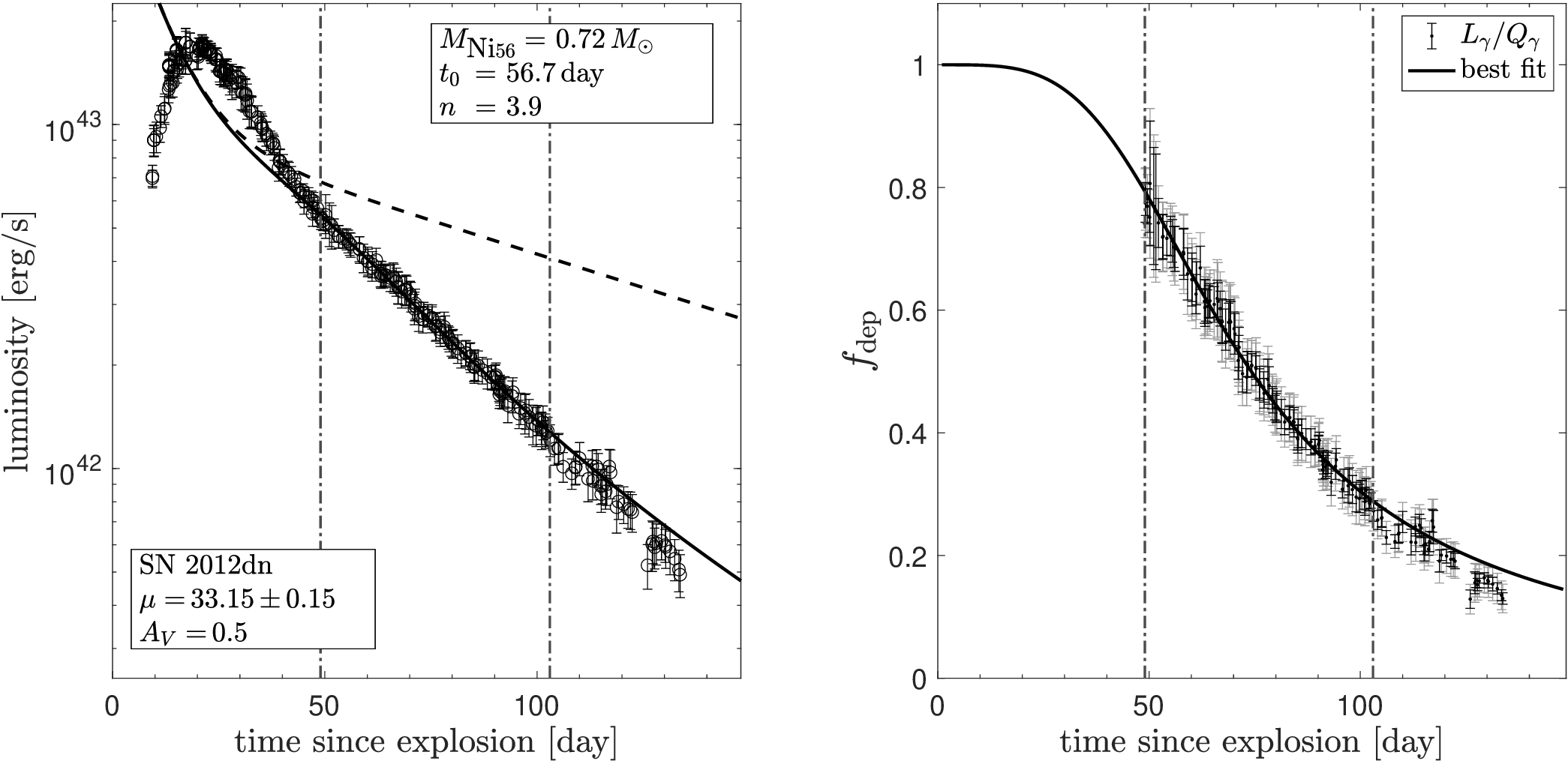}
	
	\caption{The best-fit results of the fitting method for the \fg . The median values and 68\% confidence level of the parameters are presented in Table~\ref{tab:results}. Lefthand side panels: the observed bolometric light curves are compared to the best-fit models (with parameters listed in the boxes, solid lines) and to the radioactive energy generation rates (same as  $f_{\text{dep}}=1$, dashed lines). The distance and extinction (both galactic and host) estimates are also provided in the boxes. The errors represent the total errors (both statistical and systematic). Righthand side panels: The deposition functions, $f_{\text{dep}}$, corresponding to the best-fit models (solid lines) are compared to the ratio $(L-Q_{\text{pos}})/Q_{\gamma}$. For $t \in t_{L=Q}$, this ratio corresponds to $L_{\gamma}/Q_{\gamma}$, where we use the observed $L$ and the derived $Q_{\text{pos}},Q_{\gamma}$. The total errors (statistical and systematic) are indicated by grey bars, while the statistical (photometric) errors are indicated by black bars. Vertical dashed-dotted lines mark the epochs of $t_\text{min}$ and $t_\text{max}$ in both panels ($t_\text{max}$ is not shown if when it exceeds all observations).}
	\label{fig:2009dc}
\end{figure*}

\begin{figure*}
	\ContinuedFloat
	\includegraphics[width=0.9\textwidth]{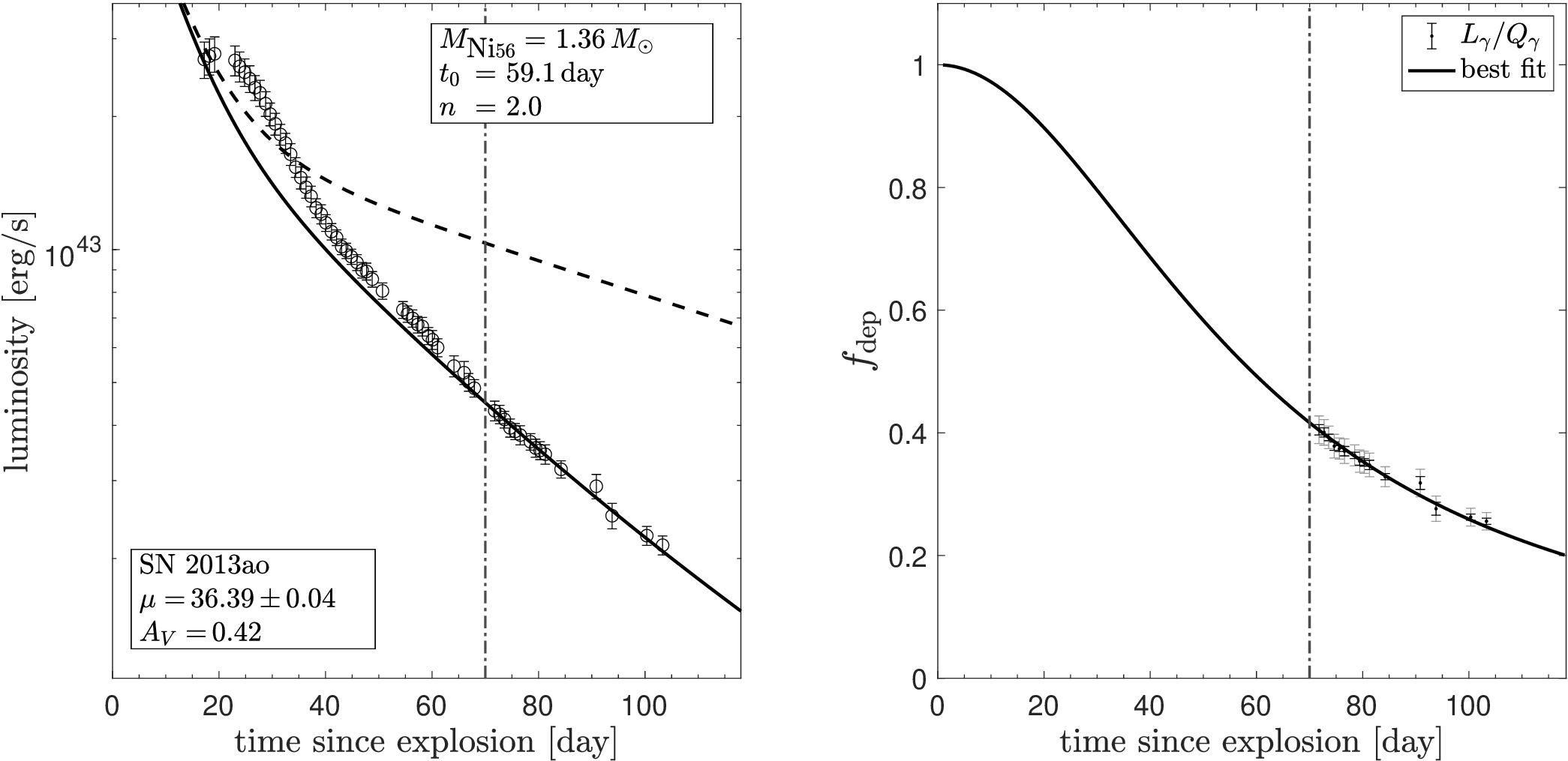}
	
	\vspace{1 cm}
	\includegraphics[width=0.9\textwidth]{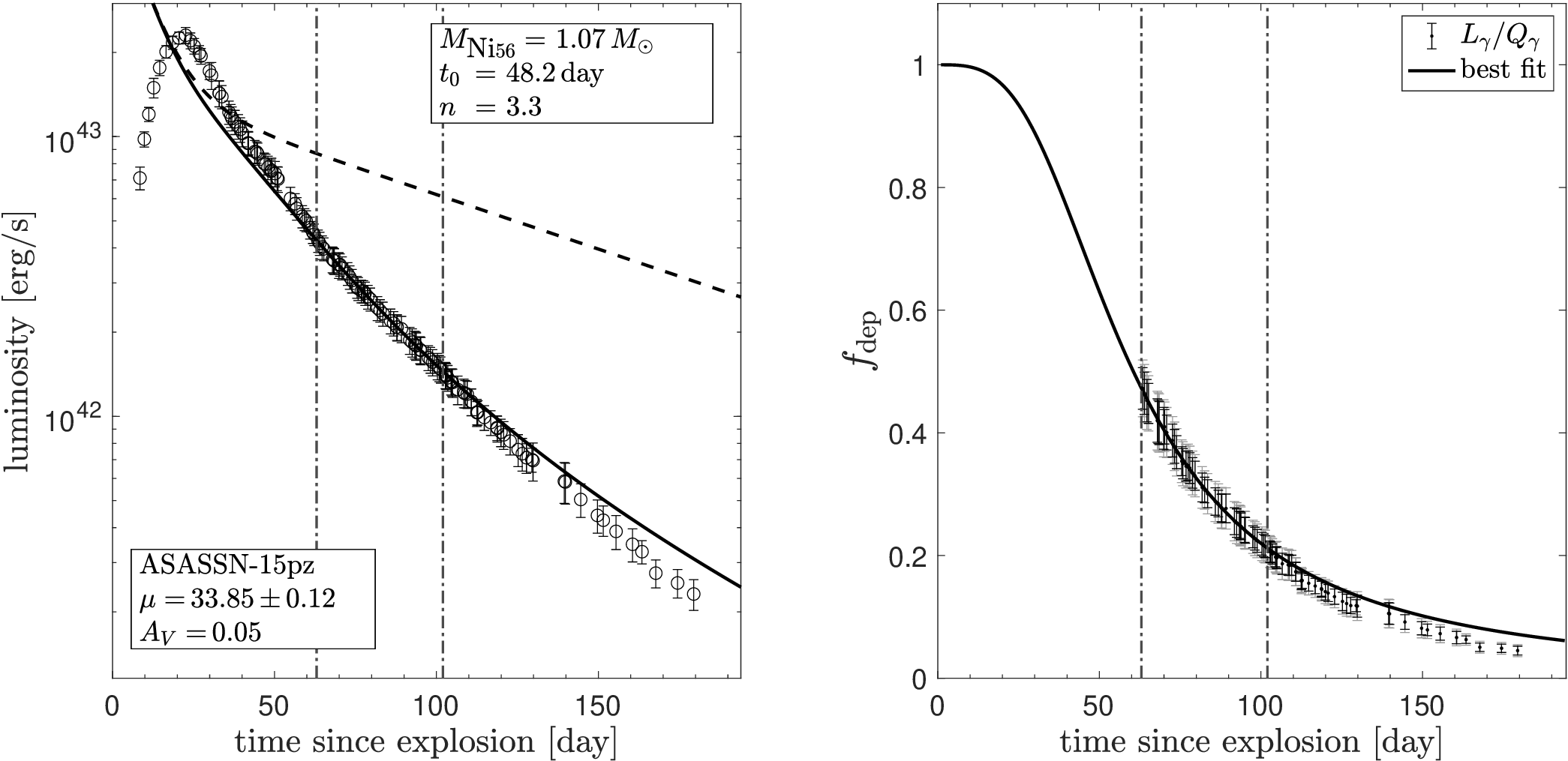}
	\caption{(continued)}
\end{figure*}

\begin{figure*}
        \ContinuedFloat
	\includegraphics[width=0.9\textwidth]{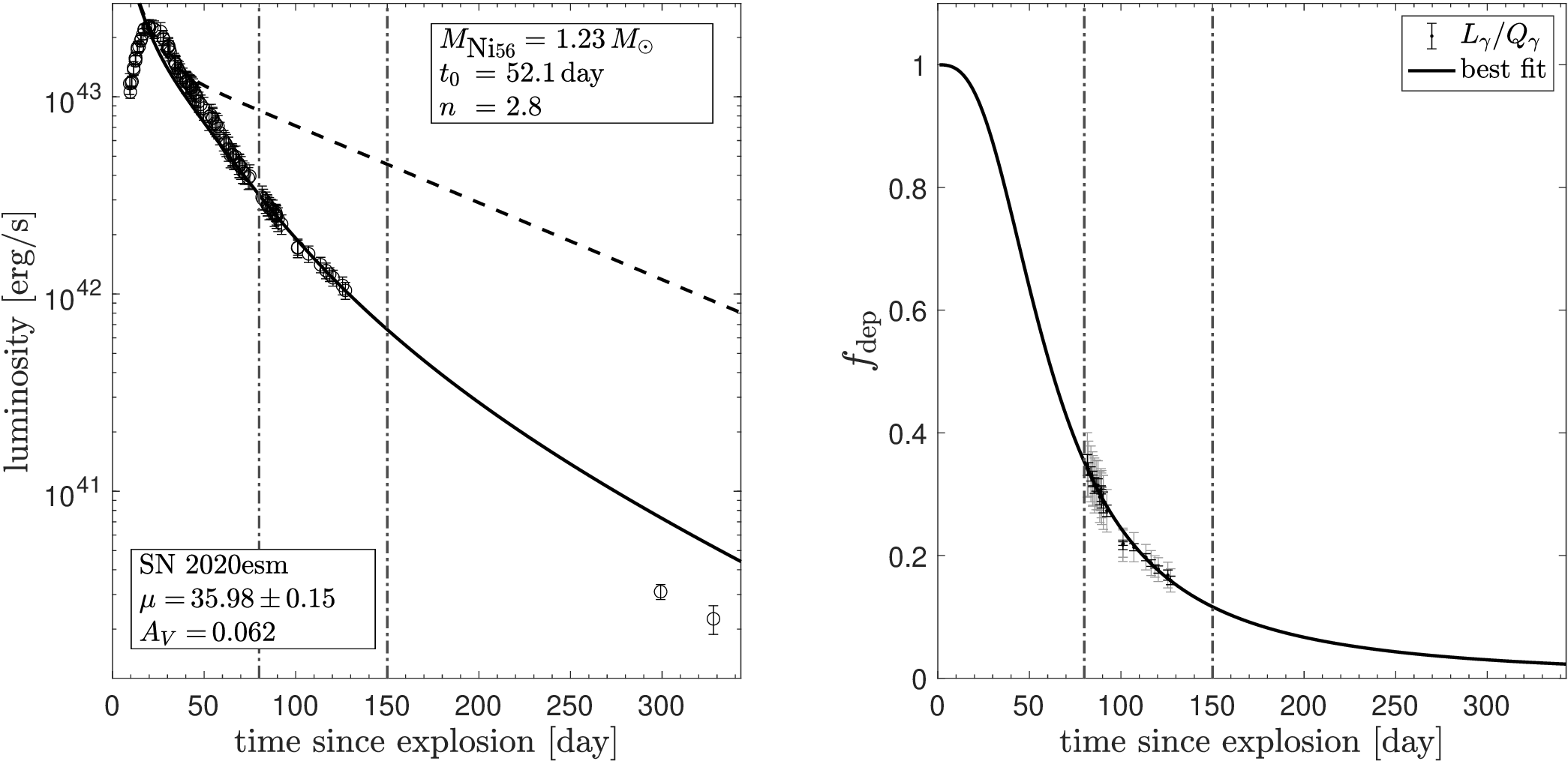}

        \vspace{1 cm}
        \includegraphics[width=0.9\textwidth]{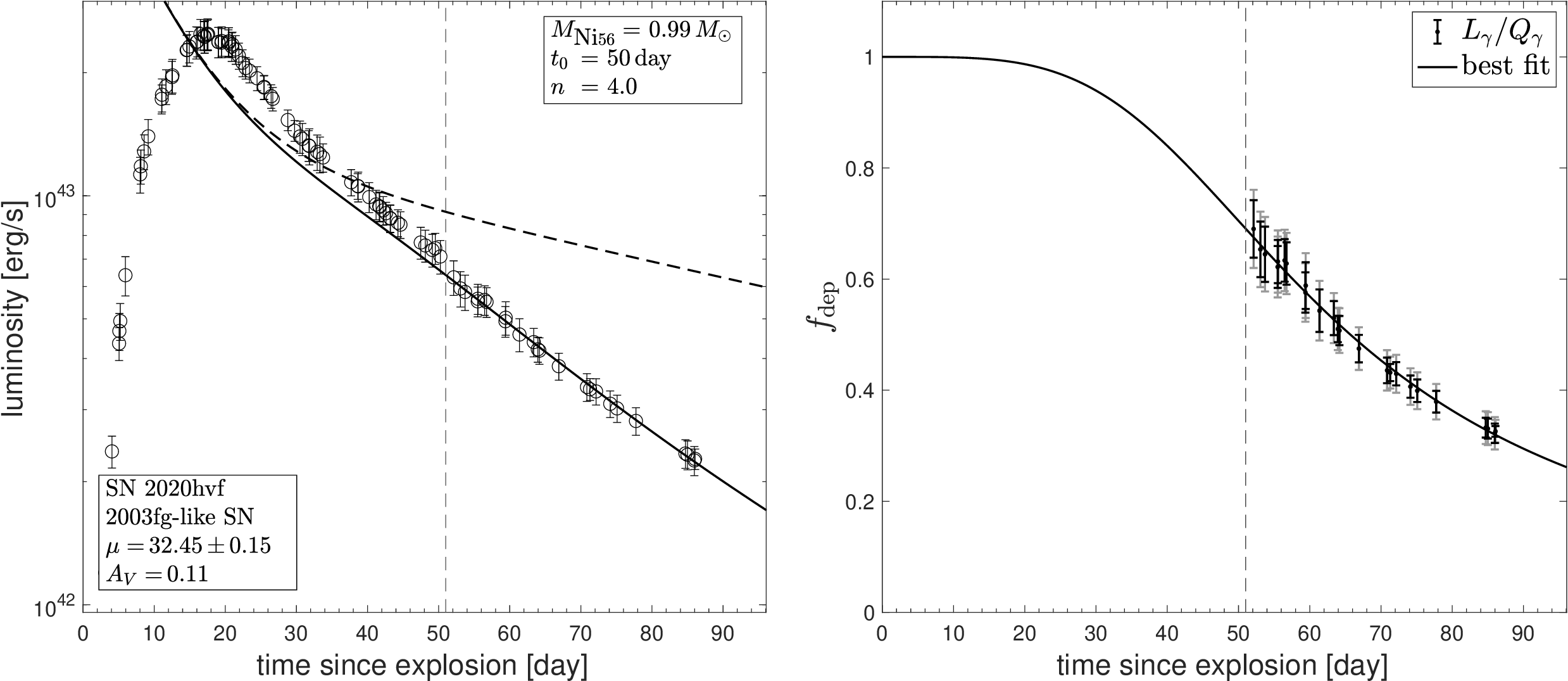}
        \caption{(continued)}
\end{figure*}

\begin{figure*}
        \ContinuedFloat	\vspace{1 cm}
	\includegraphics[width=0.9\textwidth]{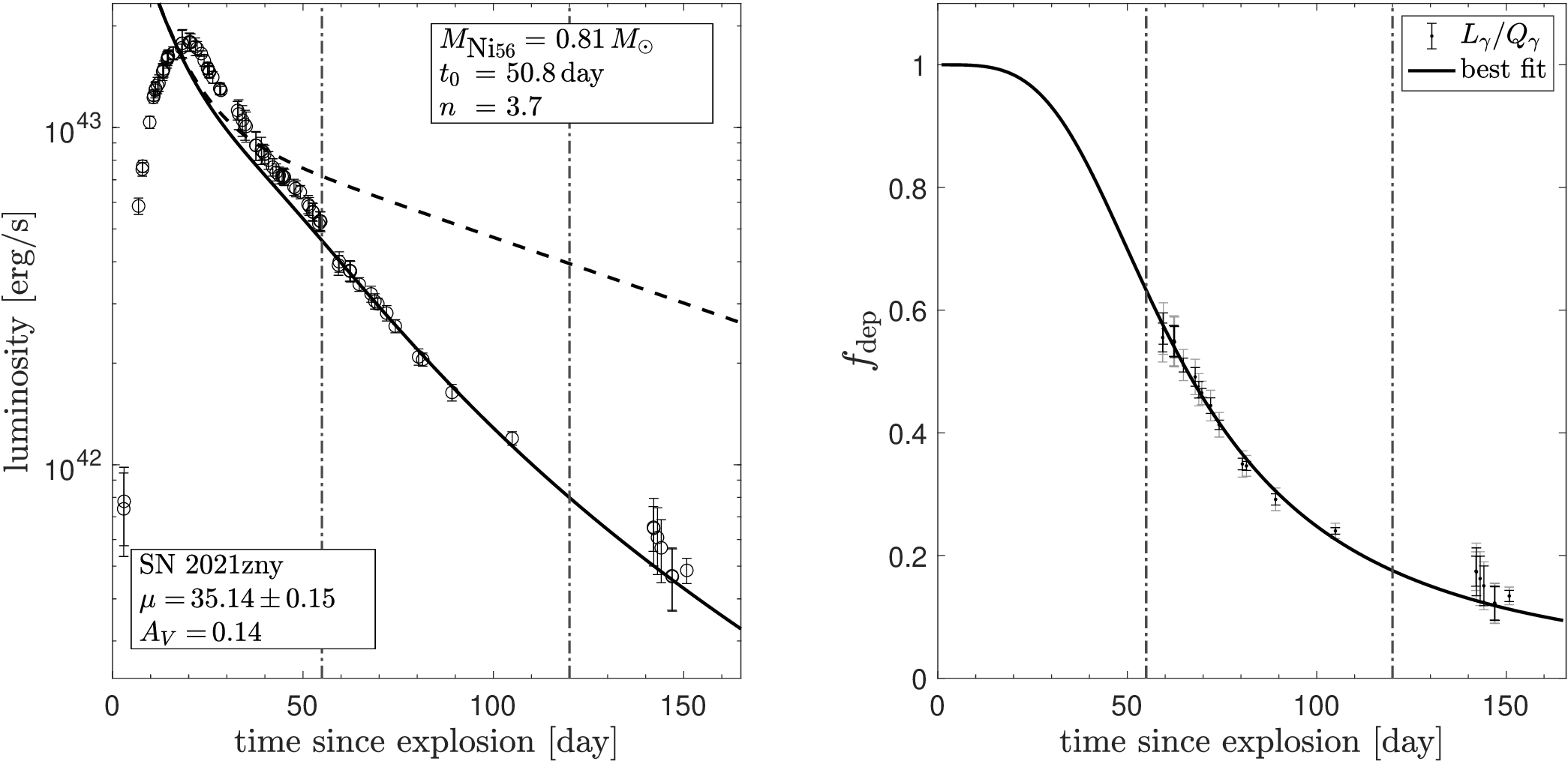}
        
        \vspace{1 cm}
	\includegraphics[width=0.9\textwidth]{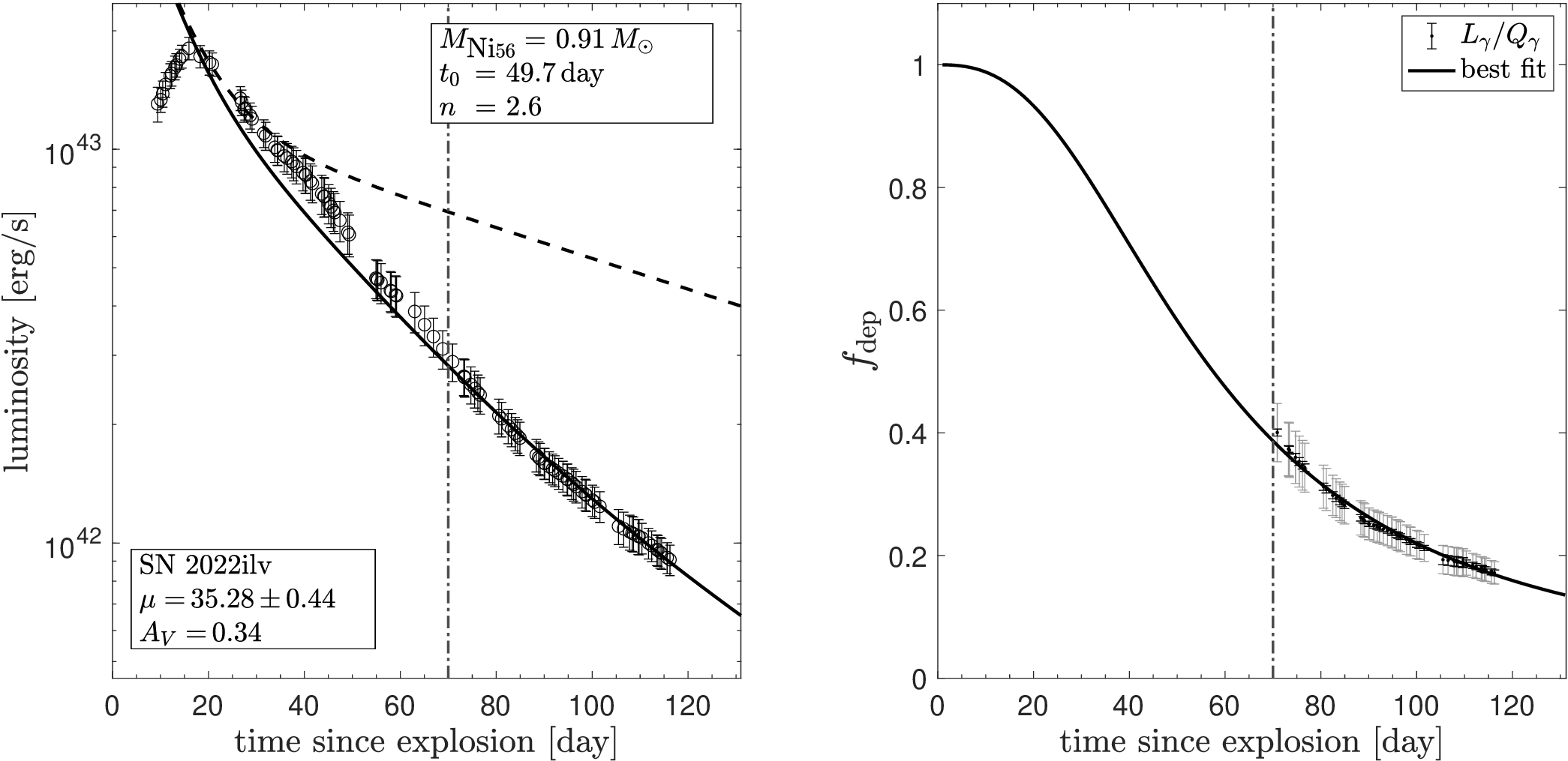}
        \caption{(continued)}
\end{figure*}

\begin{figure*}
	\includegraphics[width=0.9\textwidth]{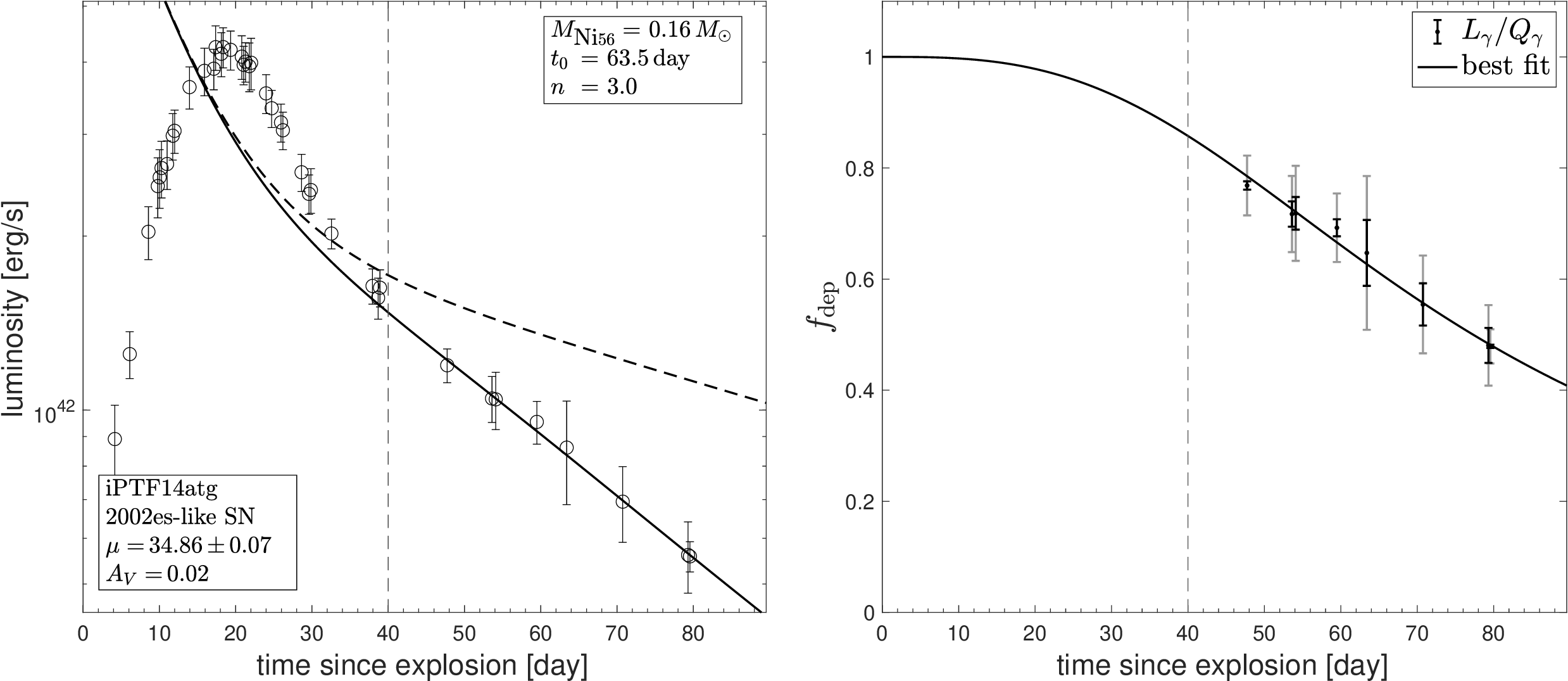}
	
	\vspace{1 cm}
	\includegraphics[width=0.9\textwidth]{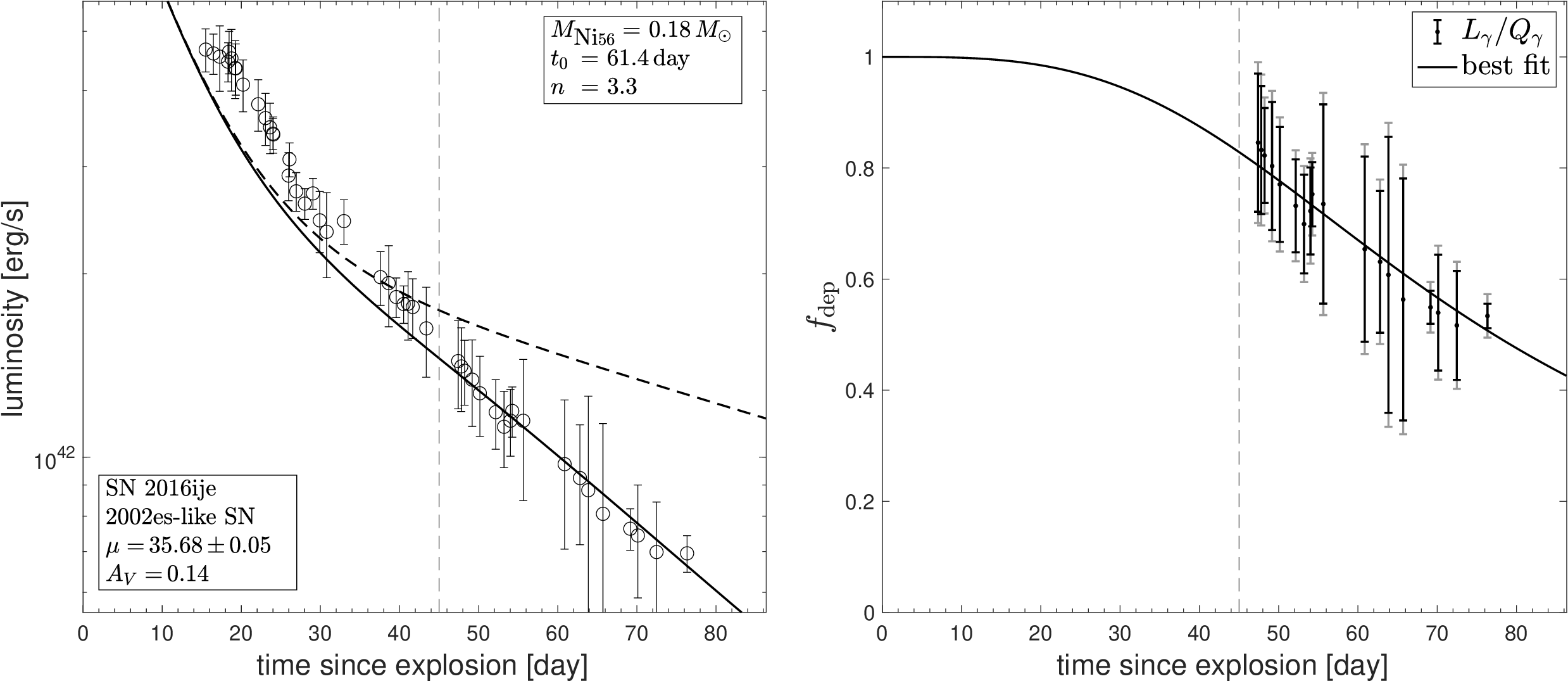}
	
	\caption{Same as Figure \ref{fig:2009dc} for the sample of \es.}
	\label{fig:fits_2002es}
\end{figure*}

\begin{figure*}
        \ContinuedFloat
	\includegraphics[width=0.9\textwidth]{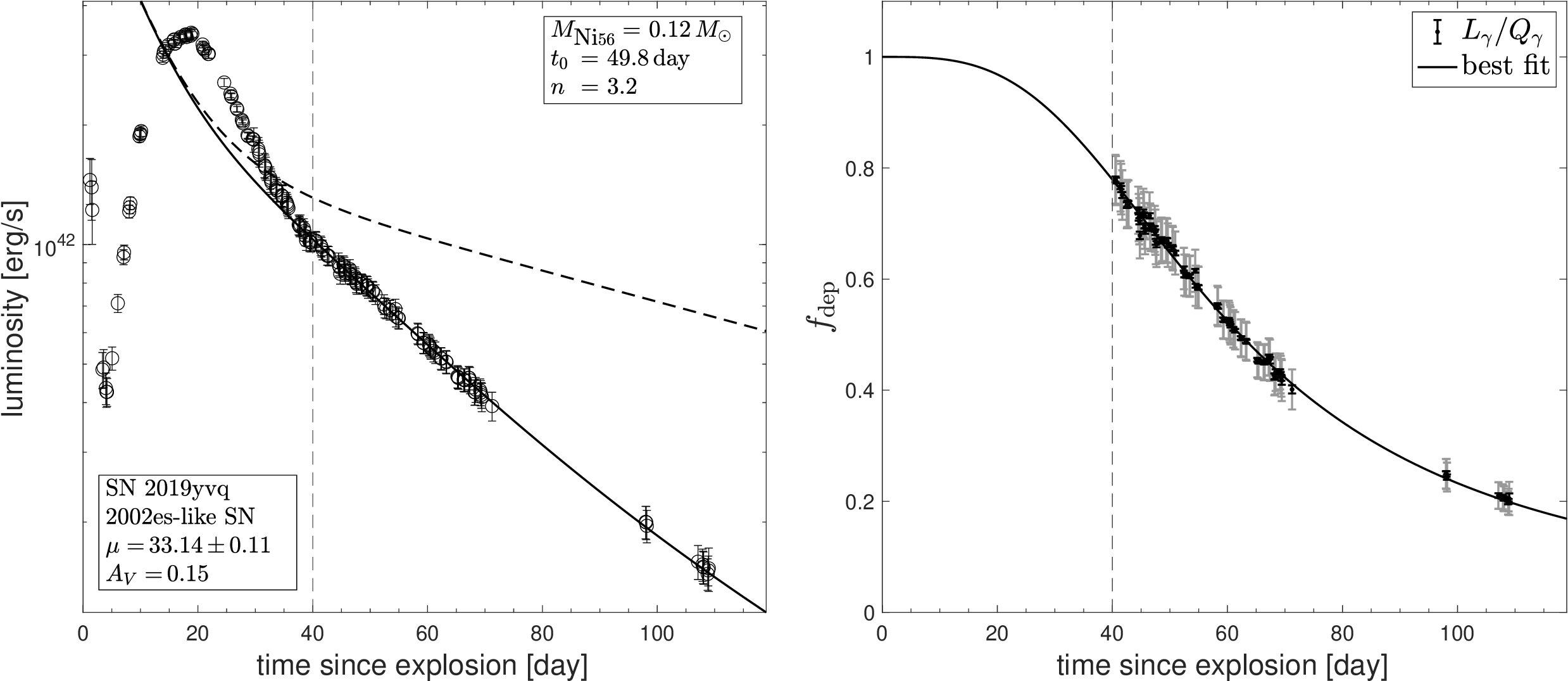}
 
        \caption{(continued)}
\end{figure*}
 
\bsp	
\label{lastpage}
\end{document}